\documentclass[12pt]{article}
\pdfoutput=1

\usepackage{jheppub} 

\usepackage[T1]{fontenc} 

\usepackage{tikz}
\usepackage{pbox}

\usepackage{amsmath}
\usepackage{booktabs}
\usepackage{multirow}

\usepackage{graphicx}

\usepackage{caption}

\usepackage{subcaption}

\usepackage{float}

\newcommand{\be}{\begin{eqnarray}}
\newcommand{\ee}{\end{eqnarray}}

\def\p{\pi}

\def\s{\sigma}

\def\s{\sigma}
\def\nn{\nonumber}
\def\p{\partial}
\def\ls{\left[}
\def\rs{\right]}
\def\lc{\left\{}
\def\rc{\right\}}

\newcommand{\bi}{\begin{itemize}}
\newcommand{\ei}{\end{itemize}}

\title{Supersymmetry Breaking and 
Inflation from  
Higher Curvature 
Supergravity}

\author[a]{I. Dalianis}
\emailAdd{dalianis@mail.ntua.gr}

\author[b]{\!\!,~F. Farakos}
\emailAdd{fotisf@mail.muni.cz}

\author[a,c]{\!\!, ~A. Kehagias}
\emailAdd{kehagias@central.ntua.gr}

\author[c]{\!\!, A. Riotto} 
\emailAdd{Antonio.Riotto@unige.ch}

\author{and  
}

\author[b]{\!\! R. von Unge}
\emailAdd{unge@physics.muni.cz}

\affiliation[a]{ Physics Division, National Technical University of Athens, \\ 15780 Zografou Campus, Athens, Greece}
\affiliation[b]{ Institute for Theoretical Physics, Masaryk University, \\
611 37 Brno, Czech Republic}
\affiliation[c]{Department of Theoretical Physics and Center for Astroparticle Physics (CAP) \\ 
24 quai E. Ansermet, CH-1211 Geneva 4, Switzerland}

\abstract{The generic embedding of the $R+R^2$ higher curvature theory into old-minimal supergravity leads to 
models with rich vacuum structure in addition to its well-known inflationary properties.  
When the model enjoys an exact R-symmetry, there is an inflationary phase with a single supersymmetric Minkowski vacuum. 
This appears to be a special case of a more generic set-up, 
which in principle may include R-symmetry violating terms which are still of pure supergravity origin. 
By including the latter terms,  
we find new supersymmetry breaking vacua 
compatible with single-field inflationary trajectories.  
We discuss explicitly two such models  and we  illustrate  how the inflaton is driven towards the supersymmetry breaking vacuum after the inflationary phase. 
In these models the gravitino mass is of the same order as the inflaton mass. 
Therefore, pure higher curvature supergravity may not only accommodate the proper inflaton field,  
but it may also provide the appropriate hidden sector for supersymmetry breaking after inflation has ended. 
 }

\begin{document} 
\maketitle
\flushbottom

\section{Introduction}

Supergravity \cite{Gates:1983nr,Wess:1992cp,Buchbinder:1998qv,Freedman:2012zz}  
provides a proper theoretical framework to describe  cosmic inflation  
in a supersymmetric set-up. 
Supersymmetry has to be  broken 
due to the absence of sparticles  in the low energy observable sector.  
Therefore, one usually explicitly introduces two matter sectors which are in general different. 
The first one is the supersymmetry breaking sector, a hidden sector serving 
only one purpose, to break supersymmetry. 
The second one is the inflationary sector which is designed to implement 
the early universe quasi de Sitter phase \cite{lr}. 
Pursuing a minimal description one can avoid the addition of a matter superfield 
that accommodates the inflaton field. 
It is possible that the inflaton has a pure gravitational origin  \cite{Starobinsky:1980te,Mukhanov:1981xt,Starobinsky:1983zz,Whitt:1984pd}, 
and thus in a supersymmetric setup, it can be a degree of freedom of a pure supergravity theory. 
Indeed, a class of higher curvature  supergravities, 
like the $R+R^2$ theory \cite{Ferrara:1978rk,Cecotti:1987sa,Cecotti:1987qe,Cecotti:1987qr}, 
incorporate supermultiplets which can accommodate the inflaton \cite{Farakos:2013cqa,FKLP}. 
In this work we show that apart from inflation, supergravity higher curvature models can implement 
at the same time the 
supersymmety breakdown.

In new-minimal supergravity the $R + R^2$ theory can be dualized to standard (old-minimal) 
supergravity coupled to a massive vector multiplet \cite{vP}, \cite{,Cecotti:1987qe,Cecotti:1987qr} 
which are solid candidates for single field inflation. 
Indeed, a single field Starobinsky model utilizing massive vector multiplets 
was proposed in  \cite{Farakos:2013cqa,FKLP}. 
This embedding has attracted further attention in a series of papers  
\cite{FKLP2,FFS1,Ferrara:2013eqa,Farakos:2014gba,Ferrara:2014cca}  due to its single field property. 
In the old-minimal supergravity, the $R+R^2$ theory is embedded  
utilizing 
two additional chiral multiplets \cite{Cecotti:1987sa}: 
the inflaton ${\cal T}$ and the (goldstino) ${\cal S}$ . 
This embedding was further developed in 
\cite{Kamada:2014gma,Kallosh:2013hoa,Ferrara:2014kva,Antoniadis:2014oya,
CK,Kallosh:2013yoa,Kallosh:2013tua,Ferrara:2013wka,
Ketov:2013dfa,Kallosh:2013lkr,Kallosh:2010xz,Farakos:2013cqa,Ferrara:2014yna,Ferrara:2014ima,Ozkan:2014cua}. 
In both embeddings, there is an inflationary phase driven by a single scalar with a potential exhibiting appropriate plateau and a supersymmetric Minkowski vacuum at the end of the inflationary phase. 
Alternative methods for embedding the Starobinsky model have been followed in  
\cite{Ketov:2014hya,Ketov:2010qz,Ketov:2012jt,Ketov:2012se,Watanabe:2013lwa,Ketov:2014qha} 
and modified no-scale supergravity models have been presented in \cite{ENO,ENO2,E2}. 
A proposal for a dynamical origin of such a phase has also been investigated in \cite{Ellis:2013zsa,Alexandre:2013nqa,Alexandre:2014lla}. 
For supergravity $N=2$ embeddings see \cite{Ketov:2014qoa,Ceresole:2014vpa,Fre:2014pca} 
and for embedding and related work in 
superstrings see \cite{Forger:1996vj,Roest:2013aoa,Kounnas:2014gda,Kitazawa:2014dya}. 
Finally for discussion on the UV properties of these models see  \cite{Kiritsis:2013gia,Copeland:2013vva}.

Although higher curvature pure supergravity may successfully accommodate the inflaton, 
as we stressed above, an extra hidden sector with matter superfields is usually introduced to break supersymmetry. 
However, the vacuum structure of the old-minimal higher derivative supergravity is rich enough \cite{Hindawi:1995qa,Hindawi:1996qi,Ferrara:2013pla}, 
and non-trivial vacua of the new degrees of freedom (associated with higher curvature gravity) may also 
lead to supersymmetry breaking 
\cite{Hindawi:1995qa,Hindawi:1996qi}. 
We find that when the inflaton ends up in an R-symmetric vacuum supersymmetry does not break \cite{Ferrara:2014cca}, 
but when the R-symmetry is violated then proper supersymmetry breaking vacua can be selected.
This gives rise to a novel approach: 
{\it the inflationary and the supersymmetry breaking sector are implemented 
by the same field configuration which is contained  in higher curvature supergravity. }
In other words, 
we investigate here the possibility that the inflationary sector and the supersymmetry breaking sector are in fact 
manifestations of pure higher curvature supergravity. 
The advantage of this approach is its universality, 
in the sense that supergravity, 
and substantially higher curvature supergravity, has to exist in every extension of the supersymmetric low energy physics that includes gravity. 
The mediation of this supersymmetry breaking to the observable sector 
requires appropriate couplings \cite{Hindawi:1996qi} 
of the higher curvature supergravity to the low energy theory. 
The details of this mediation scheme are left for a future work.

The outcome of this work is that 
inflation and supersymmetry breaking can be actually realized in a generic pure supergravity setup, 
described by 

\be
\nn
{\cal L} = -3 M_P^2 \int d^4 \theta \, E \, f({\cal R},\bar {\cal R})   
\ee 
where ${\cal R}$ is the supergravity chiral superfield which contains the Ricci scalar.   
The standard realization of the  Starobinsky model  in old-minimal supergarvity is described by 

\be
\nn
f({\cal R} , \bar {\cal R})  = 1 
- 2  \frac{ {\cal R} \bar {\cal R}}{m^2}  
+ \frac19 \zeta \, \frac{{\cal R}^2 \bar {\cal R}^2}{m^4}    
\ee 
which includes only R-symmetric terms. 
Note that this $f({\cal R} , \bar {\cal R}) $ does not lead to an $R^4$ extension as we will explicitly see below, 
and still describes $R+R^2$ supergravity. 
The $\zeta$-parameter has to 
be sufficiently large for single field inflation 
and leads to 
a unique Minkowski supersymmetric vacuum. 
Such a  model that exhibits only a unique supersymmetric 
vacuum  appears to be a rather special case. 
Here we generalize the supergravity Starobinsky model 
and investigate  superspace functions of the form 

\be
\nn
f({\cal R} , \bar {\cal R})  = 1 + \gamma \frac{{\cal R} + \bar {\cal R}}{m} 
- 2  \frac{ {\cal R} \bar {\cal R}}{m^2}  
+ \beta \frac{{\cal R} \bar {\cal R}^2 +  {\cal R}^2 \bar {\cal R}}{m^3} 
+ \frac19 \zeta \, \frac{{\cal R}^2 \bar {\cal R}^2}{m^4}    
\ee 
where $\gamma$ and $\beta$ are free coefficients which parametrize  the magnitude of the R-violation. 
In these more generic models, supersymmetry breaking vacua are present 
and the inflationary trajectory can naturally end up in  non-supersymmetric and R-violating vacua 
which can be chosen to be Minkowski as well.

The paper is organized as follows. 
In section 2, we discuss generic properties of higher curvature standard minimal supergravity, 
and the relations between the dual formulations. 
In section 3, we elaborate on the vacuum structure of the $R+R^2$ supergravity
in the old-minimal formulation and the importance of the $\zeta$ parameter  for R-symmetric models.  
Section 4 is where we turn on the $\gamma$ and $\beta$ R-symmetry violating terms. 
The theories we study in the 4th section are still of pure higher curvature supergravity origin but now 
they can play a dual role as both the inflatonary and the supersymmetry breaking sector. 
Finally we conclude in section 5.

\section{Generic pure supergravity}

Poincar\'e Supergravity, due to the 
underlying superconformal structure \cite{Lindstrom:1981qi,Gates:1983nr,Kugo:1982cu,Kugo:1983mv,Buchbinder:1998qv,Butter:2009cp,Freedman:2012zz}, 
has more than one off-shell formulations. 
In fact due to the existence of two minimal scalar multiplets which serve as compensators, 
we find Poincar\'e minimal supergravity in two flavors: 
the old-minimal \cite{Wess:1992cp}  and the new-minimal \cite{Ferrara:1988qxa}. 
In principle these formulations coincide  \cite{Ferrara:1983dh} 
when there are no curvature higher derivatives in the theory, 
but the duality can not be proven in the presence of such terms, 
and is expected to break down. 
In this work we restrict ourselves to the old-minimal formulation.

The most general (without explicit superspace higher derivatives) 
pure old-minimal  supergravity in superspace has the form 

\be
\label{LA}
{\cal L} = -3 \int d^4 \theta \, E \, f({\cal R},\bar {\cal R}) 
\ee
which can be also written as  

\be
\label{LB}
{\cal L} = \frac{3}{8} \int d^2 \Theta \,  2 {\cal E} \,  (\bar {\cal D}^2 - 8 {\cal R} )  f({\cal R},\bar {\cal R}) + c.c.
\ee
up to total derivatives. 
Here $\cal R$ is the supergravity chiral superfield and $2 {\cal E}$ is the chiral density, 
while $E$ is the full superspace density. 
For the moment we set $M_P=1$; we will restore dimensions later. 
Let us remind the reader   the superspace fields inside the Lagrangian (\ref{LB}). 
The Ricci superfield ${\cal R}$ is a chiral superfield
\be
\bar{ {\cal D}}_{\dot{\alpha}} {\cal R} =0
\ee
with lowest component the auxiliary field $M$
\be
{\cal R} | = -\frac{1}{6} M .
\ee
The fermionic component is 
\be
{\cal D}_{\alpha} {\cal R} | = 
-\frac{1}{6} ( \s^a \bar \s^b \psi_{a b} + i b^a \psi_{a } 
-i \s^a \bar \psi_{a } M )_{\alpha}
\ee
where $\psi_m^\alpha$ is the gravitino and $\psi_{mn}^\alpha$ is the gravitino field strength. 
The  highest component of this curvature chiral superfield is

\be
\begin{split}
{\cal D}^2 {\cal R} | =&  -\frac{1}{3} R + \frac{4}{9} M \bar M + \frac{2}{9} b^a b_a 
-\frac{2 i}{3} e_{a}^{\ m} {\cal D}_m b^a 
+\frac{1}{3} \bar \psi \bar \psi M -\frac{1}{3} \psi_m \s^m \bar \psi_n b^n 
\\
&+ \frac{2 i}{3} \bar \psi^m \bar \s^n \psi_{mn} 
+ \frac{1}{12} \epsilon^{k l m n} [ \bar \psi_k \bar \s_l \psi_{mn} + \psi_k \s_l \bar \psi_{mn}]  .
\end{split}
\ee
The real vector $b_m$ is an auxiliary field of the old-minimal supergravity. 
The superspace field $2 {\cal{E}}$ is a chiral density, defined as  

\be
\label{2Ep}
2{\cal{E}}=e \left\{ 1+ i\Theta \sigma^{a} \bar{\psi}_a 
-\Theta \Theta\Big{(} \bar M +\bar{\psi}_a\bar{\sigma}^{ab}\bar{\psi}_b 
\Big{)} \right\} . 
\ee
Our conventions  can be found in \cite{Wess:1992cp} 
and a full list of the components of the curvature superfields of the old-minimal supergravity 
can be found in \cite{Ferrara:1988qx}. 
The Lagrangians (\ref{LA}) and (\ref{LB}) represent a pure supergravity theory,  
and it is easy to see from the structure of the theory and the superfield ${\cal R}$ that in their generic form, 
they  lead to higher derivative gravitation.

These generic pure supergravity self-couplings can be brought in first order form by introducing appropriate Lagrange multipliers, 
giving rise to an equivalent theory which contains two chiral superfields coupled to standard supergravity, 
denoted by ${\cal T}$ and ${\cal S}$. 
It is instructive to reproduce the procedure here in the old-minimal supergravity framework. 
The Lagrangian  

\be
\label{LC}
{\cal L} = \frac{3}{8} \int d^2 \Theta \,  2 {\cal E} \,  (\bar {\cal D}^2 - 8 {\cal R} )  f({\cal S},\bar {\cal S}) + c.c. 
+6 \int d^2 \Theta \,  2 {\cal E} \, {\cal T} ({\cal S} - {\cal R} ) +c.c.
\ee
is the first step. Indeed, from the superspace equations of motion of ${\cal T}$ we have 
\be
{\cal S} = {\cal R}  
\ee
which leads to Lagrangian (\ref{LB}). 
On the other hand we may use the superspace identity

\be
\begin{split}
-6 \int d^2 \Theta \,  2 {\cal E} \, {\cal T}  {\cal R} + c.c.
&= \frac{6}{8} \int d^2 \Theta \,  2 {\cal E} \,  (\bar {\cal D}^2 -8{\cal R}) {\cal T}  +c.c.
\\
&=- 3 \int d^4 \theta \, E \, ({\cal T}  + \bar {\cal T}) 
\end{split}
\ee
to rewrite  Lagrangian (\ref{LC}) as 

\be
\begin{split}
\label{LD}
{\cal L} =& \frac{3}{8} \int d^2 \Theta \,  2 {\cal E} \,  (\bar {\cal D}^2 - 8 {\cal R} )  
\ls {\cal T}  + \bar {\cal T} + f({\cal S},\bar {\cal S})  \rs+ c.c.  \\ 
&+6 \int d^2 \Theta \,  2 {\cal E} \, {\cal T} {\cal S} +c.c.
\end{split}
\ee
which is nothing but standard supergravity

\be
\label{standard}
{\cal L} = \frac{3}{8} \int d^2 \Theta \,  2 {\cal E} \,  (\bar {\cal D}^2 - 8 {\cal R} )  
e^{- \frac{1}{3} {\cal K}}+ c.c. 
+ \int d^2 \Theta \,  2 {\cal E} \, {\cal W} +c.c.
\ee
with K\"ahler potential
\be
\label{K1}
{\cal K} = - 3 \, \text{ln} \lc  {\cal T}  + \bar {\cal T} + f({\cal S},\bar {\cal S}) \rc
\ee
and superpotential 
\be
\label{W1}
{\cal W} = 6 \, {\cal T} {\cal S} . 
\ee 
Thus,   the bottom line of the above discussion is that  the standard supergravity theory 
coupled to the chiral superfields ${\cal T}$ and ${\cal S}$ with 
a K\"ahler potential given by (\ref{K1}) and a superpotential given by  (\ref{W1}) 
is equivalent to a supergravity theory containing only  pure  supergravity sector and its higher derivatives. 
This procedure was initially established in the language of superconformal supergravity 
in \cite{Cecotti:1987sa}, and here we have followed  it closely.

The function $f({\cal R}, \bar {\cal R})$  in principle  may contain terms which violate the 
R-symmetry (here we denote them $f_{\rm v}({\cal R} , \bar {\cal R})$) and can be split as 

\be
 f({\cal R} , \bar {\cal R}) = f_{0}({\cal R} , \bar {\cal R}) + f_{\text{\rm v}}({\cal R} , \bar {\cal R})  . 
\ee
These terms have a very rich form, 
for example 
\be
f({\cal R} , \bar {\cal R})  = 1 + \gamma_n \frac{{\cal R}^n + \bar {\cal R}^n}{m^n} - 2  \frac{ {\cal R} \bar {\cal R}}{m^2}  
+ \frac19 \zeta \, \frac{{\cal R}^2 \bar {\cal R}^2}{m^4}    
\ee
where the R-symmetry violation is parameterized by $\gamma_n$, 
and leads to an equivalent theory with 
\be
{\cal K} = - 3 \, \text{ln} \lc 1   +  {\cal T}  + \bar {\cal T} 
+ \gamma_n \frac{{\cal S}^n  + \bar {\cal S}^n}{ m^n} 
- 2  \frac{{\cal S} \bar {\cal S}}{m^2} 
+ \frac19 \zeta \, \frac{{\cal S}^2 \bar {\cal S}^2}{m^4}  \rc    
\ee
and   the superpotential is given by (\ref{W1}). 
A more generic R-symmetry breaking can originate from 

\be
 f_{\text{\rm v}}({\cal R} , \bar {\cal R}) \supset {\cal R}^p  \bar {\cal R}^q 
\ee
for $p \ne q$. 
Let us also mention another useful equivalence. 
As we saw the function $f$ can be split such that it has the form 

\be
\label{split}
f({\cal R},\bar {\cal R}) = h({\cal R}) + \bar h(\bar {\cal R})  + g({\cal R},\bar {\cal R}) . 
\ee
From our previous discussion it is easy to see that the higher derivative supergravity theory 
in (\ref{split}) is equivalent to standard supergravity with superpotential (\ref{W1}) and K\"ahler potential 

\be
\label{K2}
{\cal K} = - 3 \, \text{ln} \lc  {\cal T}  + \bar {\cal T} +  h({\cal S}) + \bar h(\bar {\cal S})  + g({\cal S},\bar {\cal S}) \rc . 
\ee
But there is another equivalent form this theory may have. 
By shifting 
\be
{\cal T} \rightarrow {\cal T} -  h({\cal S}) 
\ee
the K\"ahler potential (\ref{K2}) becomes (\ref{K1}) (with $f({\cal S},\bar {\cal S})$ 
replaced by $g({\cal S},\bar {\cal S})$) and the superpotential now reads  
\be
\label{W2}
{\cal W} = 6 \, {\cal T} {\cal S} - 6 \, {\cal S} \, h({\cal S}) . 
\ee 
This equivalence has been discussed for the supergravity sector in  \cite{Hindawi:1995qa} 
and also later found to exist in the superconformal supergravity \cite{Ferrara:2013wka}. 
Note that if we set  \cite{Lindstrom:1979kq,Farakos:2013ih,Antoniadis:2014oya,Ferrara:2014kva}  
\be
{\cal S} = X_{NL}
\ee
where
\be
\label{S2}
X_{NL}^2=0 
\ee
the most general K\"ahler potential reads 

\be
\label{S2S}
{\cal K} = - 3 \, \text{ln} \lc  {\cal T}  + \bar {\cal T} + \gamma X_{NL} 
+ \gamma \bar X_{NL} -2  X_{NL} \bar X_{NL} \rc  
\ee
but in  fact due to the aforementioned equivalence the linear terms $ \gamma X_{NL}$ and $\gamma \bar X_{NL}$  
inside (\ref{S2S}) are redundant 
and vanish. 
As we will see later, these linear terms play a fundamental role in our work, 
therefore the use of a constrained superfield (\ref{S2}) is not appropriate, 
and will have to be modified for our models.

For future reference we write down the bosonic sector of a standard supergravity theory 
with K\"ahler potential (\ref{K1}) and superpotential (\ref{W1})

\be
\label{Lcomp}
e^{-1} {\cal L} = -\frac12 R - {\cal K}_{ij} \p z^i \p \bar z^j - {\cal V}(z^i,\bar z^i)  
\ee
where we denote collectively the complex scalar fields $T$ (the lowest component  of ${\cal T}$)  
and $S$ (the lowest component  of ${\cal S}$) as $z^i$ 
and the potential has the standard form

\be
\label{sugrapot}
{\cal V} = e^{\cal K} \ls ({\cal K}^{-1})^{i \bar j} ({\cal W}_i + {\cal K}_i {\cal W})(\bar {\cal W}_{\bar j} + {\cal K}_{\bar j} \bar {\cal W}) - 3 {\cal W} \bar {\cal W}  \rs . 
\ee
In formula (\ref{Lcomp}) we have 

\be
{\cal K}_{ij} \p z^i \p \bar z^j = -3 \frac{f_{S \bar S} \p S \p \bar S}{(f + T + \bar T)} 
+ 3 \frac{|f_S \p S + \p T|^2}{(f + T + \bar T)^2} 
\ee
and
\be
\!\!\!\!\!\!
{\cal V} = 12 \frac{|S|^2 \Big{(} f - 2 (f_S S + f_{\bar S} \bar S) +4 f_{S \bar S} |S|^2 \Big{)} 
- (f_{S \bar S})^{-1} |T -f_S S + 2 f_{S \bar S} |S|^2 |^2}{(f + T + \bar T)^2}. 
\ee
When restoring the canonical dimensions for the fields of the previous formulae 
one merely has to compensate them with appropriate $M_P$ powers, 
since the new scales are hidden inside the function $f$.

Finally we would like to clarify 
the relevance of the model (\ref{LA}) 

\be
{\cal L} = -3 \int d^4 \theta \, E \, f({\cal R},\bar {\cal R})  
\ee
in describing $R+R^2$ gravity. 
Writing down the bosonic sector of theory in the pure supergravity picture one finds  \cite{Hindawi:1995qa} 
\be
\label{FR1}
\begin{split}
{\cal L} =& - \frac12 \left( f + M f_M + \bar M f_{\bar M} - 4 M \bar M f_{M \bar M} 
- 2 b^m b_m f_{M \bar M} \right) R - \frac34  f_{M \bar M} R^2
\\
&  + 3  f_{M \bar M} \, \p M \p \bar M  - 3  f_{M \bar M} \, (\nabla^m b_m)^2  
+ i \left( f_M \p^m M - f_{\bar M} \p^m \bar M  \right) b_m 
\\
& - i \left( f_M M - f_{\bar M} \bar M \right) \nabla^m b_m 
- \frac13 M \bar M \lc f - 2 \left( f_M M + f_{\bar M} \bar M \right) + 4 M \bar M f_{M \bar M} \rc 
\\
& + \frac13 b_m b^m \lc f  +  f_M M + f_{\bar M} \bar M  - 4 M \bar M f_{M \bar M} - b_n b^n f_{M \bar M} \rc  
\end{split}
\ee 
where $f = f\left( -\frac16 M, -\frac16 \bar M \right) $, $f_M=\p f / \p M$ and $f_{\bar M}=\p f / \p {\bar M}$. 
A   very interesting discussion on (\ref{FR1}) can be found in  \cite{Hindawi:1995qa}. 
Note that this is a theory of  curvature and curvature square terms only, 
as far as gravitation is concerned. 
Nevertheless it does not propagate the same degrees of freedom as standard gravity. 
The $R+R^2$ theory on top of the dynamical  degrees of freedom of the metric, 
also gives rise to an additional real scalar propagating degree of freedom \cite{Whitt:1984pd} known as the scalaron. 
For a gravitational theory the counting of the degrees of freedom ends here.  
For the supergravitational embedding, as can be seen from (\ref{FR1}), the scalaron comes with 
supersymmetric scalar partners.  
The counting of the total scalar supersymmetric degrees of 
freedom is \cite{Ferrara:1978rk,Antoniadis:2014oya,Hindawi:1995qa} 
\begin{itemize}

\item $M$: 2 real scalar degrees of freedom, 

\item $\nabla^m b_m$: 1 real scalar degree of freedom,  

\item $R^2 \rightarrow scalaron$: 1 real scalar degree of freedom.

\end{itemize} 
The above degrees of freedom reside inside appropriate supersymmetric multiplets as shown in \cite{Ferrara:1978rk}, 
and are the reason why one needs exactly two chiral superfields (${\cal S}$ and ${\cal T}$) 
to perform the duality of the theory (\ref{FR1}) 
to standard supergravity; the degrees of freedom should match.

We stress that even though in the following sections we will work with the models in the dual picture, 
one may always turn to the pure supergravity description by the use of formula (\ref{FR1}), 
but the on-shell properties of the two descriptions and the dynamics will be exactly the same, 
leading to identical predictions.

The situation during inflation is quite subtle. 
When the field $M$ is strongly stabilized to $\langle M \rangle = 0$  one finds the Starobinsky model.  
But when $M$ is not strongly stabilized, 
there will be a slight translation of the vev of $M$ during inflation. 
The dynamics of these models is in general highly involved but they are related to a well known class of gravitational theories. 
In fact the situation is similar to \cite{Kehagias:2013mya}. 
If we start from (\ref{FR1}) it is easy to see that for small kinematic terms of $M$ 

\be
\label{slowM}
3  f_{M \bar M} \, \p M \p \bar M \ll 
- \frac13 M \bar M \lc f - 2 \left( f_M M + f_{\bar M} \bar M \right) + 4 M \bar M f_{M \bar M} \rc 
\ee
the field $M$ becomes an auxiliary field leading to algebraic equations of motion. 
Thus in the limit (\ref{slowM}) the equations for $M$ will have solutions of form  
\be
M = {\cal M} (R)
\ee
and after plugging back to the Lagrangian, this theory will be nothing but an ${\cal F}(R)$ gravity 
\be
\label{FR2}
e^{-1} {\cal L} = {\cal F}( R)  . 
\ee
To summarize: 
\begin{enumerate}
\itemsep-.1em 
\item For vanishing 
$\langle M \rangle 
$, the inflationary phase is described by the Starobinsky model.
\item For non-constant  $\langle M \rangle
$,  
the inflationary phase is described by a general ${\cal F}(R)$ model.
\end{enumerate}
Thus, even thought this is a theory of quadratic gravitation its inflationary dynamics has various phases 
depending on the choice of $f({\cal R},\bar {\cal R})$. 
As we will see, it also has a rich vacuum structure.

\section{$ R + R^2$  in old-minimal supergravity and the $\zeta$ parameter}

There are two different directions towards the embedding of the Starobinsky model of inflation \cite{Starobinsky:1980te} 
in  old-minimal supergravity in a superspace setup. 
One is essentially the embedding of the inflationary potential 
\be
\label{Vstar}
{\cal V} = \frac{3}{2} m^2 M_P^2  ( 1- e^{ - \sqrt \frac23 \varphi/M_P} )^2 
\ee
for the real scalar inflaton field $\varphi$, 
by employing various supemultiplets.  
The second method is somewhat more geometrical, 
and it consists of embedding the higher derivative model 
\be
\label{Rstar}
e^{-1} {\cal L} = - \frac12 M_P^2 R + \frac{M_P^2}{24 m^2} R^2 
\ee
into supergravity, 
thus one employes a pure supergravitational sector \cite{Farakos:2013cqa}. 
Due to the well-known duality \cite{Whitt:1984pd} between 
a gravitational theory coupled to a propagating scalar with potential 
given by (\ref{Vstar}) and a higher derivative gravitation of the form (\ref{Rstar}) 
(which we reproduced in a supergravity framework in the previous section), 
the two aforementioned methods of embedding the Starobinsky model in supergravity lead to equivalent results. 
Note that for Starobinsky inflation the new scale $M$ is in general taken as  
\be
m \sim  10^{-5} M_P .
\ee

The $R+R^2$ supergravity was initially found in the linearized level in \cite{Ferrara:1978rk}. 
Later the embedding  was extended to the 
full theory of the old-minimal supergravity \cite{Cecotti:1987sa}, 
and it corresponds to the  following superspace Lagrangian 

\be
\label{LE}
{\cal L} = -3 M_P^2 \int d^4 \theta \, E \, \ls 1  - 2  \frac{{\cal R} \bar {\cal R}}{m^2} \rs  
\ee
or in other words to a specific function $f({\cal R},\bar {\cal R})$ inside (\ref{LA}) or (\ref{LB}) 

\be
\label{fRR}
f({\cal R},\bar {\cal R}) = 1 - 2  \frac{{\cal R} \bar {\cal R}}{m^2}  
\ee
where we have now restored the Planck mass $M_P$ for the rest of this article. 
Taking into account our discussion in the previous section, 
we see that the equivalent standard supergravity system coupled to matter is described 
by a  K\"ahler potential

\be
\label{K22}
{\cal K} = - 3 M_P^2 \, \text{ln} \lc 1 + \frac{{\cal T}  + \bar {\cal T}}{ M_P}   - 2  \frac{{\cal S} \bar {\cal S}}{M_P^2} \rc
\ee
and superpotential 
\be
\label{W2}
{\cal W} = 6 m \, {\cal T} {\cal S}  
\ee 
where following \cite{Hindawi:1995qa}  we have redefined 
\be
{\cal S} \rightarrow \frac{m}{M_P} {\cal S} . 
\ee
Indeed for the extremum 
\be
\label{st0}
\langle S \rangle = \langle \text{Im} T \rangle = 0 
\ee
the potential for the canonical normalized inflaton, 
which is the appropriately redefined $\text{Re} T$,  is given by (\ref{Vstar}).

In \cite{Kallosh:2013lkr} it was pointed out that for the specific K\"ahler potential (\ref{K22}) 
and superpotential (\ref{W2}) the extremum (\ref{st0}) 
during inflation is  unstable due to a tachyonic mass for $S$. 
This instability was proposed to be remedied by 
introducing an additional term in the K\"ahler potential 
which includes the $\zeta$ parameter 

\be
{\cal K} = - 3 M_P^2 \, \text{ln} \lc 1 +  \frac{{\cal T}  + \bar {\cal T}}{ M_P}  
- 2  \frac{{\cal S} \bar {\cal S}}{M_P^2} 
+ \frac19 \zeta \, \frac{{\cal S}^2 \bar {\cal S}^2}{M_P^4}  \rc . 
\ee
This has a pure supergravity origin in the dual picture since it corresponds to setting  

\be
\label{fR4}
 f({\cal R} , \bar {\cal R}) = 1  - 2  \frac{ {\cal R} \bar {\cal R}}{m^2}  
+ \frac19 \zeta \, \frac{{\cal R}^2 \bar {\cal R}^2}{m^4}  
\ee
inside (\ref{LA}) or (\ref{LB}).

The $\zeta$ parameter controls the vacuum structure of the models, 
and there are some specific critical values, which lead to  different properties. 
We wish to discuss this further. 
The scalar potential is

\be 
\label{VV}
\begin{split}
{\cal V} =& 12 m^2 M_P^2 \Big{(} 1 - 2 (s^2 + c^2 ) + \frac{\zeta}{9} (s^2 + c^2 )^2 + 2 t  \Big{)}^{-2} \times 
\\
& \Big{[} (s^2 + c^2) \lc1 - 2(s^2+c^2) + \zeta (s^2 + c^2)^2  \rc  
\\
& + \frac{9}{2} \frac{[t - \frac23 (s^2 +c^2) (3 - \zeta (s^2 +c^2)) ]^2 + b^2 }{(9 - 2 \zeta (s^2 +c^2 ))}  \Big{]}  
\end{split}
\ee
where we have set
\be
\label{TS}
\begin{split}
\frac{T}{M_P} &=  t + i b 
\\
\frac{S}{M_P} &=  s + i c .
\end{split}
\ee
Note that 
\be
\frac{\p^2 {\cal V}}{\p s^2} \Big{|}_{b=s=c=0} = m^2 \lc \frac{2916 t^2}{(9+18 t)^3} 
+ \frac{81 (4-8t)}{2 (9+18 t )^2} + \frac{18 t^2 \zeta}{(9+18 t)^2}  \rc 
\ee
which sets a lower bound for the value of $\zeta$ such that the mass of the $s$ field is  
not tachyonic \cite{Kallosh:2013lkr}. 
In our conventions we find
\be
\label{klbound}
\zeta  
\gtrsim 3.54 . 
\ee

In the remainder of this section we further study the structure of the new, 
supersymmetry breaking  vacua.  
As far as their inflationary properties are concerned, 
only the models with $\zeta > 3.54$ are solid candidates since otherwise the scalar $S$ may lead to instabilities. 
The model enjoys a rich vacuum structure due to the term 
\be
 \frac19 \zeta \, \frac{{\cal R}^2 \bar {\cal R}^2}{m^4}  
\ee
but an inflationary phase exists only for (\ref{klbound}). 
We will show in the next section that turning on R-symmetry violating terms new vacua emerge 
and also new inflationary trajectories. 
For now let us return to the vacuum structure of (\ref{VV}).

\begin{figure}
\centering{
\includegraphics[width=0.79\textwidth]{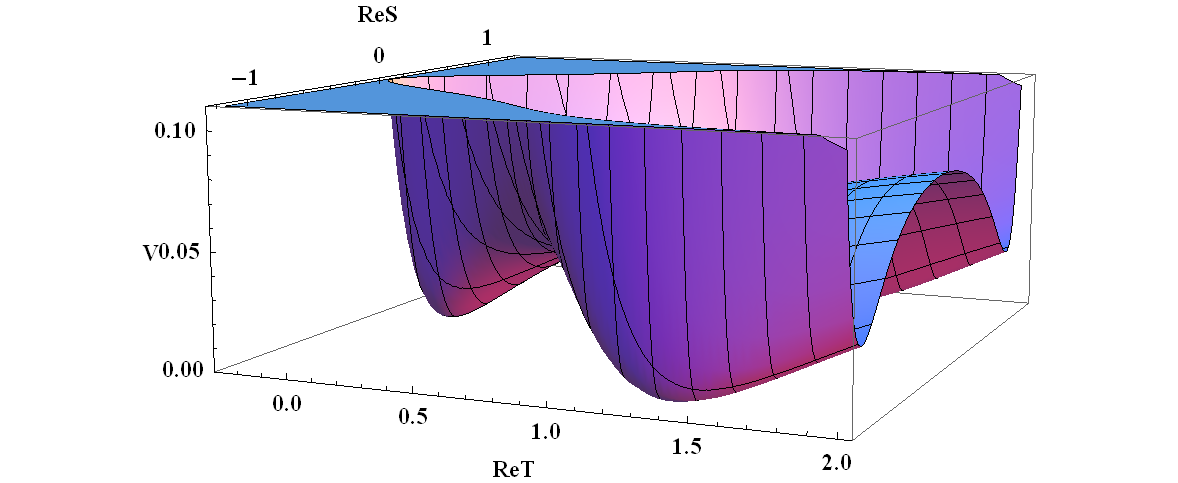}}
\caption{The shape of the scalar potential (\ref{VVMINK}) in units of $12 m^2 M_P^2$. 
The scalar fields $T$ and $S$ are in $M_P$ units.  }
\label{PLVVMINK}
\end{figure}

We shortly  review the work of   \cite{Hindawi:1995qa} where 
\be  
\zeta = 1 . 
\ee 
We use the parametrization (\ref{TS}) for $T$ 
and the full potential reads   \cite{Hindawi:1995qa} 

\be 
\label{VVMINK}
\begin{split}
{\cal V} =& 12 m^2 M_P^2 \Big{(} 1 - 2  \frac{|S|^2}{M_P^2} 
+ \frac{1}{9} \frac{|S|^4}{M_P^4} + 2 t  \Big{)}^{-2} \times 
\\
& \ls  \frac{|S|^2}{M_P^2} \lc1 - 2 \frac{|S|^2}{M_P^2} +   \frac{|S|^4}{M_P^4} \rc  
 + \frac{9}{2} \frac{\ls t - \frac23  \frac{|S|^2}{M_P^2} 
\left(3 -   \frac{|S|^2}{M_P^2}\right) \rs^2 +  b^2 }{\left(9 - 2   \frac{|S|^2}{M_P^2} \right)}  \rs .    
\end{split}
\ee
It is easy to see that  
\be
\langle b \rangle = 0 
\ee
in the vacuum state and since the potential depends only on $|S|^2 $ 
it has  a flat direction along the argument of $S$. 
A plot of the potential (\ref{VVMINK}) with 
$\langle \text{Im} S \rangle = \langle \text{Im} T \rangle = 0$ can be found in Fig. \ref{PLVVMINK}. 
Following the method in \cite{Hindawi:1995qa} we find the following vacua 
\begin{itemize}
\itemsep-.1em 
\item 
Supersymmetric Minkowski vacuum: $\langle S \rangle = \langle T \rangle=0$

\item 
Mikowski vacuum with broken supersymmetry: $\langle |S| \rangle = M_P$, $ \langle T \rangle= \frac43 M_P $ . 

\end{itemize}
In the supersymmetry breaking vacuum the R-symmetry is broken spontaneously and the flat direction is 
parameterized by the argument of $S$ which is nothing else but the R-axion. 
These vacua were originally found in \cite{Hindawi:1995qa} . 

\begin{figure}
\centering
\begin{subfigure}{.5\textwidth}
  \centering
 \includegraphics[width=1.17\linewidth]{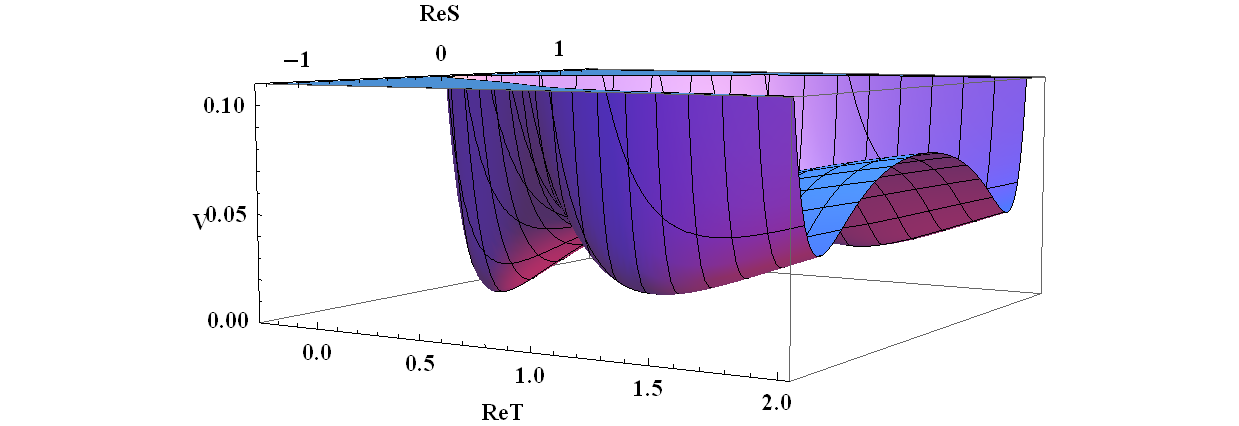}
\caption{ }
\label{PLVVDS}
\end{subfigure}%
\begin{subfigure}{.5\textwidth}
  \centering
 \includegraphics[width=0.81\linewidth]{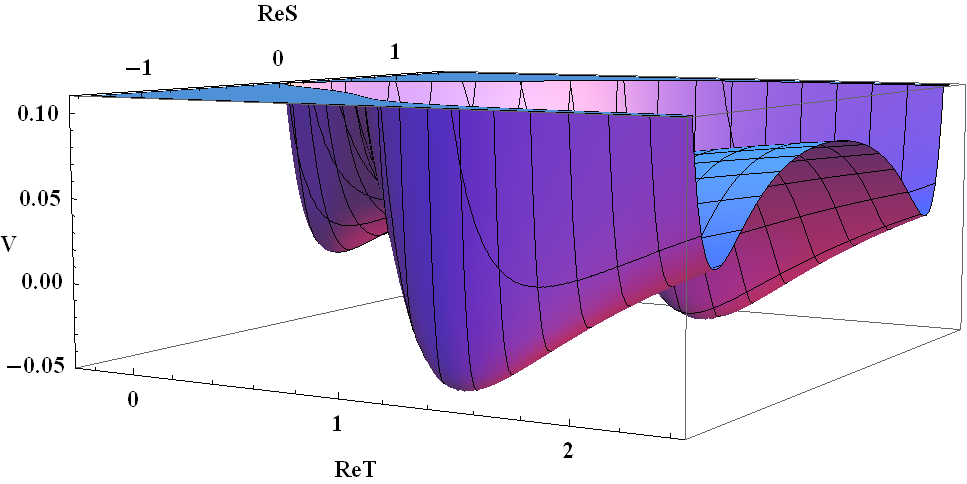} 
\caption{ }
\label{PLVVADS}
\end{subfigure}
\caption{The shape of the scalar potential (\ref{VV}) for meta-stable de Sitter vacua with $\zeta=1.1$ (left panel) 
and anti-de Sitter vacua with $\zeta=0.9$ (right panel) in 
units of $12 m^2 M_P^2$. 
 The scalar fields $T$ and $S$ are in $M_P$ units.}
\label{fig:test}
\end{figure}

We can further develop this setup and investigate 
the properties of the models with  small deviations of $\zeta$ from the special value $\zeta=1$ . 
We are not in contrast with the work of \cite{Kallosh:2014oja} since here we deform a Minkowski vacuum 
with broken supersymmetry. 
If we set 
\be
\zeta \gtrsim 1
\ee 
it is easy to understand that the SUSY-breaking Minkowski vacua  will get slightly uplifted 
leading to a metastable de Sitter vacuum. 
For 
$\zeta \lesssim 1$
the SUSY-breaking vacua will get slightly dragged down.  
This is illustrated in the plots in Fig. \ref{PLVVDS} and Fig. \ref{PLVVADS}. 
Note that in the supersymmetry breaking vacuum the spontaneous breakdown of the R-symmetry 
gives rise to an R-axion. 
Metastable supersymmetry breaking vacua have also been encountered in the past in a different setup, 
as for example in \cite{Dalianis:2010yk,Dalianis:2011zz}.

We can summarize the dependence of the vacuum structure of the R-symmetric models 
(for the relevant field space region) 
on the various $\zeta$ values 
\begin{itemize}

\itemsep-.04cm 

\item $\zeta > 3.54 \  \text{(Kallosh-Linde  bound \cite{Kallosh:2013lkr})}$: 
The $S$ field is always stabilized at $\langle S \rangle =0 $ 
and the model has a single supersymmetric Minkowski vacuum for $\langle T \rangle =0$.  
For larger values of $\zeta$  the  mass of the $S$-field 
increases and becomes bigger than the Hubble scale during inflation, 
therefore does not influence the dynamics.

\item $  \zeta \gtrsim  1$: The model has two  minima. 
The  Minkowski one with $\langle S \rangle =\langle T \rangle =0$ which preserves supersymmetry and the local 
de Sitter one with $\langle |S|^2 \rangle \sim M_P^2 $  and $\langle T \rangle \sim  M_P$, 
which breaks both supersymmetry and R-symmetry. 

\item $\zeta = 1 \  \text{(Hindawi-Ovrut-Waldram critical value \cite{Hindawi:1995qa})}$: 
The model has two degenerate Minkowski local minima. 
The trivial $\langle S \rangle =\langle T \rangle =0$ which preserves supersymmetry and the 
new one with $\langle |S|^2 \rangle = M_P^2 $  and $\langle T \rangle = \frac43 M_P$, 
which breaks both supersymmetry and R-symmetry with vanishing cosmological constant. 

\vskip -1in

\item $ \zeta \lesssim 1 $: The model has two minima. 
The trivial Minkowski one $\langle S \rangle =\langle T \rangle =0$ which preserves supersymmetry and the 
global anti-de Sitter one with $\langle |S|^2 \rangle \sim M_P^2 $  and $\langle T \rangle \sim  M_P$, 
which breaks both supersymmetry and R-symmetry. 

\item $ \zeta =0 \  \text{(Cecotti critical value \cite{Cecotti:1987sa})}$: The model has only one stable vacuum, 
the trivial Minkowski one which preserves supersymmetry. For $\zeta < 0 $ the model will in principle suffer from instabilities. 

\end{itemize}
The above results have been collected in the following  table \ref{table}.

%
%

\begin{table}[H]
\centering 
\begin{tabular}{|| l || p{0.15\textwidth} | p{0.14\textwidth} | p{0.18\textwidth} | l ||} 
\hline\hline 
$\zeta$-parameter & Minkowski  & de Sitter  & anti-de Sitter & 
\pbox{2cm}{tachyonic \\ instability }  
\\ [0.5ex]
\hline\hline
$\zeta > 3.54$ & \checkmark(+) &   &  &  no
\\[0.5ex]
\hline
$  \zeta  \gtrsim 1 $ & \checkmark(+) & \checkmark$(-)$ &  & yes
\\[0.5ex]
\hline
$\zeta = 1$ & $\checkmark(+), ~\checkmark(-)$    &  & &yes
\\[0.5ex]
\hline
$\zeta \lesssim  1   $ & \checkmark(+) &  & \checkmark$(-)$ & yes
\\[0.5ex]
\hline
$\zeta =  0 $ & \checkmark(+) &   &  & yes
\\[.003ex]
\hline
\hline 
\end{tabular}
\caption{R-Symmetric models. 
The $\checkmark(+)$ symbol stands for supersymmetric vacua, 
while $\checkmark(-)$ for vacua with broken supersymmetry.}  
\label{table}
\end{table}

\section{Including R-symmetry violating terms}

In the previous section we saw higher curvature supergravity may have Minkowski vacua with broken supersymmetry, 
but this led to two drawbacks. 
Firstly, for a Minkowski vacuum one has to set $\zeta =1$ and therefore 
the same model can not describe a single field inflationary phase. 
Secondly, these vacua led to the spontaneous breaking of the R-symmetry, giving rise to the massless R-axion. 
Even for a de Sitter vacuum with a small cosmological constant these problems persist. 
In this section we introduce R-symmetry violating terms and we show how single field inflation 
and supersymmetry breaking can be realized in this class of models 
avoiding at the same time the R-axion massless mode.

\subsection{Vacuum structure for $f_ {\rm v} = \frac{\gamma}{M_P} \left( {\cal S} + \bar {\cal S} \right)$ }

Our first model is given by 

\be
\label{RRRR}
f ({\cal R} , \bar {\cal R})= 1 + \gamma \, \frac{ {\cal R}}{m} + \gamma \, \frac{ \bar {\cal R}}{m} 
- 2  \frac{ {\cal R} \bar {\cal R}}{m^2}  
+ \frac19 \zeta \, \frac{{\cal R}^2 \bar {\cal R}^2}{m^4}  
\ee
where the second and third term source the R-symmetry breaking, 
and the equivalent K\"ahler potential  is 

\be
{\cal K} = - 3 M_P^2 \, \text{ln} \lc 1   +  \frac{{\cal T}  + \bar {\cal T}}{ M_P}  
+  \gamma \frac{{\cal S}  + \bar {\cal S}}{ M_P} 
- 2  \frac{{\cal S} \bar {\cal S}}{M_P^2} 
+ \frac19 \zeta \, \frac{{\cal S}^2 \bar {\cal S}^2}{M_P^4}  \rc  
\ee
while the superpotential is 
\be
{\cal W} = 6 m \, {\cal T} \, {\cal S} . 
\ee
The scalar potential for this model reads
\be 
\label{VV6}
\begin{split}
{\cal V}(t, b, s, c) =& 12 m^2 M_P^2 \Big{(} 1 - 2 (s^2 + c^2 ) + \frac{\zeta}{9} (s^2 + c^2 )^2 + 2 t + 2 \gamma s \Big{)}^{-2} \times 
\\
& \Big{[} (s^2 + c^2) \lc1 - 2(s^2+c^2) + \zeta (s^2 + c^2)^2 -2 \gamma s \rc  
\\
&  + \frac{9}{2} \frac{[t - \gamma s - \frac23 (s^2 +c^2) (3 
- \zeta (s^2 +c^2)) ]^2 +  (\gamma c + b)^2 }{(9 - 2 \zeta (s^2 +c^2 ))}  \Big{]} 
\end{split}
\ee
in the parametrization (\ref{TS}). 
This potential will in general have two classes of vacuum solutions.  
First there is the trivial vacuum 
\be
\label{v0}
\langle  T \rangle = \langle  S \rangle = 0 
\ee
with no supersymmetry breaking and vanishing vacuum energy. 
Then there is the new class of vacua 
\be
\label{vBR}
\begin{split}
\langle  T \rangle &= M_P \langle  t \rangle + i M_P \langle b \rangle = M_P \, t_0  
\\
\langle  S \rangle &=  M_P \langle  s \rangle + i M_P \langle c \rangle   = M_P \, s_0  
\end{split}
\ee
which will break supersymmetry with vanishing vacuum energy 
or small positive (or negative) vacuum energy depending on the parameters $\zeta$ and $\gamma$.

To find the non-trivial vacuum we follow a similar procedure as in \cite{Hindawi:1995qa}. 
Let us first explain why  the imaginary components of $T$ and $S$ are stabilized to the origin
\be
\langle c \rangle = \langle b \rangle = 0 .  
\ee
It is easy to see from the term 
$ (\gamma c + b)^2 $
inside the potential (\ref{VV6}) that 
\be
\label{thisterm}
\langle b \rangle = - \gamma \langle c \rangle . 
\ee
When (\ref{thisterm}) is satisfied the $c$-field appears  quadratically in (\ref{VV}) and does not have 
tachyonic mass, hence it will be stabilized at 
$ \langle c \rangle = 0$ . 
Then the potential becomes

\be 
\label{VVapp}
\begin{split}
{\cal V}(t, s) =& 12 m^2 M_P^2 \Big{(} 1 - 2 s^2 + \frac{\zeta}{9} s^4 + 2 t + 2 \gamma s \Big{)}^{-2} \times 
\\
& \Big{[} s^2  \lc1 - 2 s^2 + \zeta s^4 -2 \gamma s \rc 
+ \frac{9}{2} \frac{[t -  \gamma s -\frac23 s^2 (3 - \zeta s^2) ]^2  }{(9 - 2 \zeta s^2 ) }  \Big{]}  .  
\end{split}
\ee
Since the denominator is positive-definite, 
the model will have local Minkowski vacua around the field space regions 
where the numerator is non-negative: 

\be
\label{app1}
s^2  \lc1 - 2 s^2 + \zeta s^4 -2 \gamma s \rc 
+ \frac{9}{2} \frac{[t -  \gamma s - \frac23 s^2 (3 - \zeta s^2) ]^2  }{(9 - 2 \zeta s^2 ) } \ge 0 . 
\ee
For the regions where 
$9 - 2 \zeta s^2 > 0$   
it is easy to see from the second term in (\ref{app1})
that the vacuum expectation value for $t$ will be at 
\be
t_0 = \gamma s_0 + \frac23 s_0^2 (3 - \zeta s_0^2) . 
\ee
Now we look for the regions of the $s$ field space where
the first term of  (\ref{app1}) 

\be
\omega_{\gamma}(s) = 1 - 2 s^2 + \zeta s^4 -2 \gamma s 
\ee
has a local minimum $s_0$ with 
\be
\label{eqapp1}
\omega_{\gamma}(s_0) = 0 . 
\ee
The requirement  for  local minimum implies 
\be
\label{eqapp1B}
\frac{\p \omega_\gamma(s)}{\p s}\Big{|}_{s_0} &=& 0  
\\
\label{eqapp1C}
\frac{\p^2 \omega_\gamma(s)}{\p s^2}\Big{|}_{s_0} &>& 0 . 
\ee
The above expressions (\ref{eqapp1}) and (\ref{eqapp1B}) 
 relate $s_0$ to $\zeta$ and $\gamma$ as follows 

\be
\zeta &=& \frac{1 + 2 s_0^2}{3 s_0^4} 
\\
\gamma &=& -2 s_0 + \frac{2 + 4 s_0^2}{3 s_0}  
\ee
where we remind the reader that $s_0 = \langle s \rangle$. 
Thus, the parameters $\zeta$ and $\gamma$ for Minkowski vacua 
are parameterized by the vacuum expectation value of the $s$-field, 
which should be chosen such that 

\be
\label{app3}
9 -  \frac{2 + 4 s_0^2}{3 s_0^2}  > 0    
\ee
according to second term in (\ref{app1}) and should also realize (\ref{eqapp1C}). 
We conclude that there is a whole parameter space for $\zeta(s_0)$ and $\gamma(s_0)$ which 
give Minkowski vacua with broken supersymmetry.

Let us now turn to the mass of the gravitino  field at the vaccum which reads 

\be
\label{grmass}
m^2_{3/2} = \langle e^{\cal K} |{\cal W}|^2 \rangle = \frac{ 24  m^2 s_0^2 (1+2s_0^2)^2}{11-14 s_0^2}  .  
\ee 
The above expression yields  a  bound on the possible values of $s_0$ 

\be
\label{bon}
|s_0| < \sqrt \frac{11}{14} 
\ee
such that $m^2_{3/2} > 0 $. 
Interestingly, this bounds $\langle S \rangle$ to sub-Planckian values. 
Moreover (\ref{bon}) also gives $\gamma > 0.16 $ and $\zeta> 1.38 $.  
It is well known that whenever  the gravitino acquires a non-vanishing mass in 
a Minkowski vacuum 
\be
\label{vacuumrelation}
\langle ({\cal K}^{-1})^{i \bar j} (D_i {\cal W})(D_{\bar j} \bar {\cal W}) \rangle 
- 3 \langle {\cal W} \bar {\cal W} \rangle = 0 
\ee
this signals supersymmetry breaking 
\be
\langle D_i {\cal W} \rangle \ne 0 . 
\ee
Note that even though supersymmetry is broken and the vacuum energy is vanishing the gravitino 
mass is not specified unless a value for the parameter $s_0$ is chosen.

\begin{figure}[tbp] 
\centering{
\includegraphics[scale=0.35]{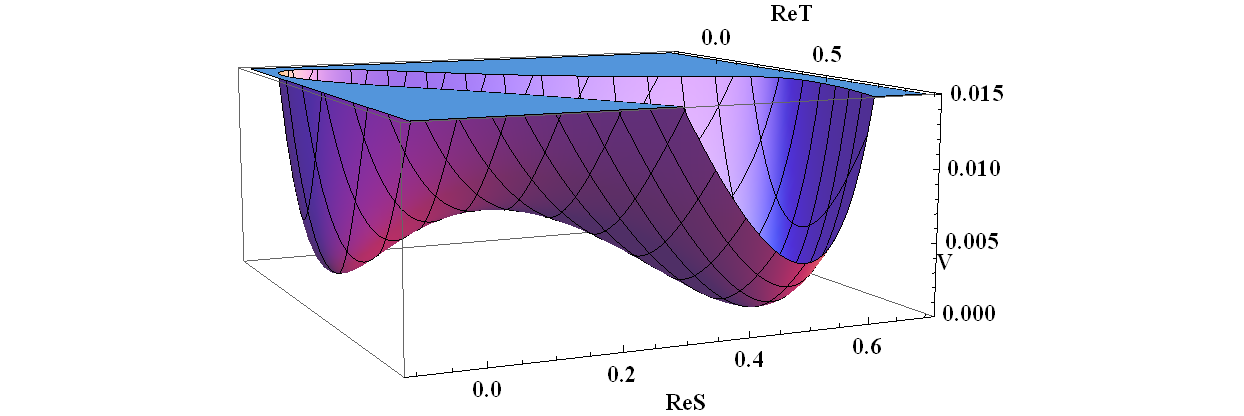}}
\caption{The shape of the scalar potential (\ref{VV6}) with $\zeta=8$ and $\gamma=1$ in units of $12 m^2 M_P^2$. 
The scalar fields $T$ and $S$ are in $M_P$ units.  }
\label{linSVV}
\end{figure}

We illustrate the features of the vacuum structure in a simple example where  
\be
\zeta = 8  
\ \ , \ \ 
\gamma = 1 . 
\ee
By following our previous discussion it is easy to see that the potential (\ref{VV6}) will have a 
supersymmetry preserving vacuum (\ref{v0}) and on top of that there will be the vacuum 

\be
\begin{split}
\langle  T \rangle &= M_P t_0 =  \frac{2}{3} M_P  
\\
\langle  S \rangle &= M_P s_0   =  \frac{1}{2} M_P  .  
\end{split}
\ee
The shape of  the scalar potential (\ref{VV6}) for $\zeta=8$ and $\gamma=1$  
can be seen in Fig. \ref{linSVV} where it is easy to identify the two  minima with vanishing vacuum energy 
and subplanckian vacuum expectation values. 
An investigation of the kinematic terms gives 

\be
\left( f(S, \bar S) + \frac{T  }{M_P} + \frac{\bar T}{  M_P} \right) f_{S \bar S} < 0  
\ee 
which shows that they are positive definite around the second vacuum. 
Supersymmetry is broken and  gravitino mass is found to be

\be
m^2_{3/2} = \langle e^{\cal K} |{\cal W}|^2 \rangle \simeq \frac{7}{4} \, m^2   
\ee 
while $\langle D_T {\cal W} \rangle \approx \langle D_S {\cal W} \rangle \approx  \, m \, M_P$.

Finally we should note that for $\zeta=8$ and  
\be
\gamma \gtrsim 1
\ee
the model describes metastable de Sitter vacua, 
while for 
\be
\gamma \lesssim 1
\ee
anti-de Sitter local minima are found.

Here we briefly study  the global limit of the model about the supersymmetry breaking vacua. 
Let us have $\zeta = 8$,  $\gamma = 1$.  
Our method to find the global model around the second SUSY-breaking vacuum 
is to expand the K\"ahler potential and superpotential in powers of $M_P$ 
as proposed for the similar case in \cite{Hindawi:1995qa}, 
even though alternative decoupling limits exist \cite{deAlwis:2012aa}. 
To expand around the second vacuum  we set

\be
{\cal T} &=& \frac{2}{3} M_P + \tilde {\cal T}
\\
{\cal S} &=& \frac{1}{2} M_P + \tilde {\cal S} . 
\ee
After diagonalizing the fields by setting 
$\tilde {\cal X} = \tilde {\cal T} - \frac79  \tilde {\cal S}$ we have a global supersymmetric model with 

\be
{\cal K} &=& 3 {\tilde {\cal X}} {\bar{\tilde {\cal X}}} + \frac{10}{27} {\tilde {\cal S}} {\bar { \tilde {\cal S}}} 
\\
{\cal {\cal W}} &=& 3 \mu^2 \tilde {\cal X} + \frac 13 \mu^2  \tilde {\cal S}  
\ee
where $\mu = \sqrt{ m M_P}$. 
The supersymmetry breaking leads to

\be
\langle {\cal V} \rangle =  \frac{33}{10} \mu^4  .  
\ee

\subsection{Inflationary properties of $f_{\rm v} =  \frac{\gamma}{M_P} \left( {\cal S} + \bar {\cal S} \right)$}

As we have mentioned the above models describe higher derivative supergravity and   
due to the existence of more than one scalar fields the vacuum structure is rich. 
In this subsection we show that there are directions in the field space that can successfully drive inflation.  
This class of models, apart from R-symmetric terms,  includes the R-violating terms controlled by  the parameter  

\be
\gamma \ne 0 . 
\ee

The models contain two complex scalars, $T$ and $S$ where their imaginary parts are stabilized to 
$\text{Im} T = b = 0$ and $\text{Im} S = c = 0$.  
The relevant inflationary scalar potential reads 

\be 
\label{VVinf}
\begin{split}
{\cal V}(t, s) =& 12 m^2 M_P^2 \Big{(} 1 - 2 s^2 + \frac{\zeta}{9} s^4 + 2 t + 2 \gamma s \Big{)}^{-2} \times 
\\
& \Big{[} s^2  \lc1 - 2 s^2 + \zeta s^4 -2 \gamma s \rc 
+ \frac{9}{2} \frac{[t -  \gamma s -\frac23 s^2 (3 - \zeta s^2) ]^2  }{(9 - 2 \zeta s^2 ) }  \Big{]}  .  
\end{split}
\ee
For a Minkowski vacuum with broken supersymmetry one asks for

\be
\zeta \rightarrow \zeta_0 = \frac{1 + 2 s_0^2}{3 s_0^4} \ \ , \ \  
\gamma \rightarrow \gamma_0 = -2 s_0 + \frac{2 + 4 s_0^2}{3 s_0}  . 
\ee
Here $s_0$ is the vacuum expectation value of $\text{Re}S$ which serves as a free parameter for the Minkowski vacua.

\subsubsection{Inflationary trajectory}

\begin{figure}[tbp] 
\centering{
\includegraphics[scale=0.5]{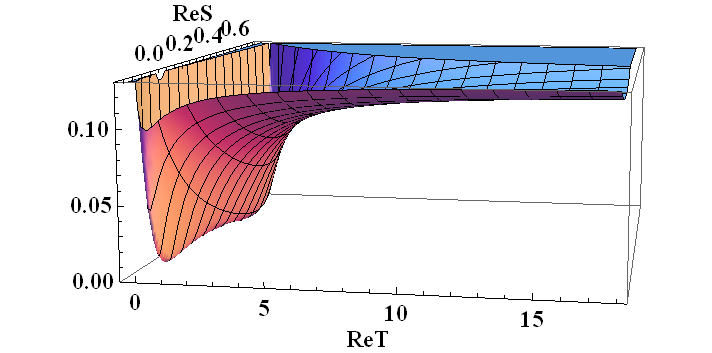}}
\caption{The shape of the scalar potential (\ref{VV6}) with $\zeta=8$ and $\gamma=1$ in units of $12 m^2 M_P^2$. 
The scalar fields $T$ and $S$ are in $M_P$ units. A plateau similar to the Starobinsky model can be seen for large Re$T$ values. 
The inflationary trajectory terminates at the SUSY breaking vacuum where Re$S = 0.5$.} 
\label{linSVV12}
\end{figure}

\begin{figure}
\centering
\begin{subfigure}{.5\textwidth}
  \centering
  \includegraphics[width=.99\linewidth]{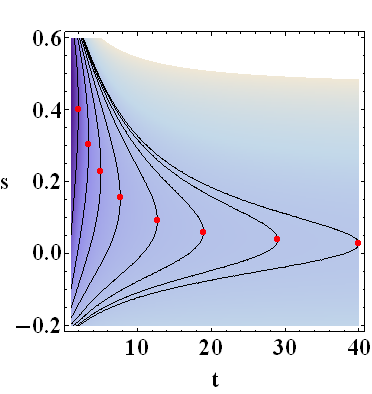}
  \caption{}
  \label{PLVVinf}
\end{subfigure}%
\begin{subfigure}{.5\textwidth}
  \centering
  \includegraphics[width=.99\linewidth]{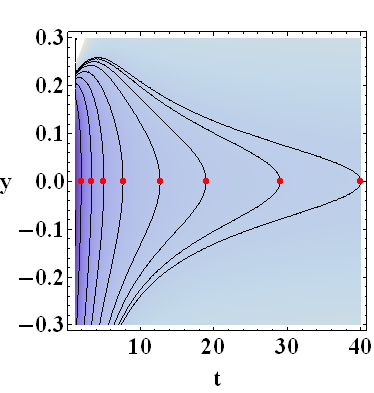}
  \caption{}
  \label{PLVVxt}
\end{subfigure}
\caption{The trajectory in field space Re$T$ - Re$S$ for $\gamma=1$ and $\zeta=8$ 
for the original fields in the scalar potential (\ref{VVinf}) (left) 
and for the redefined fields in the scalar potential (\ref{VVinfQt}) (right). }
\label{fig:test}
\end{figure}

This model inflates for large $t$ values during which $s$ takes small values close to zero. 
The $s$ field constantly drifts away from zero, 
and approaches $s_0$ at the end of inflation. 
This can be seen at the plot  in Fig. \ref{PLVVinf} for the scalar potential ${\cal V}(t, s)$.  
An important feature is that the field configuration is attracted to the broken vacuum, 
thus the model will inflate and then settle down to the supersymmetry breaking vacuum.  
A plot can be seen in Fig. \ref{linSVV12}.

Although there  are two fields $t$ and $s$ which  are varying 
we will show that there is a field redefinition which reduces the system to a single field case. 
The procedure to find the proper fields is straightforward. 
To achieve this  we define a new field which is a combination of the original ones, 
and  is stabilized at zero 

\be
\langle y(s,t) \rangle = 0 . 
\ee
Note that the field $s$ varies very slowly and 
it will always be stabilized at the minimum of its potential for the given $t$ thanks to the large $\zeta$ value. 
Hence the value of $s$ can be always parameterized by the value of $t$.

By calculating the first derivative of the potential with respect to $s$ we have  

\be
\label{dVds}
\frac{\p {\cal V}(t,s)}{\p s} =0 
\ee
which is the $s$-direction minimum. 
Equation (\ref{dVds}) is algebraic and  can in principle be solved for $s$ in terms of $t$, that is  

\be
s = {\cal Q}(t)  
\ee
leading to 

\be
\label{Qt}
\frac{\p {\cal V}(t,s)}{\p s} \Big{|}_{s = {\cal Q}(t)  }=0 . 
\ee
For the potential (\ref{VVinf}) an expansion of the equation (\ref{dVds}) is 

\be
\label{e1}
t  = a_0 \frac1s + a_1 +a_2 s +a_3 s^2  + {\cal O}(s^3) 
\ee
where 

\be
\begin{split}
a_0 &= \frac{1296 \gamma}{144 \zeta}
\\
a_1 &= - \frac{36(\zeta -36)}{144 \zeta}
\\
a_2 &= - \frac{36+288 \gamma^2 - \zeta}{144 \gamma} 
\\
a_3 & = \frac{8 \gamma^2 (\zeta - 99) + 9(4 - \zeta)}{144\gamma^2} . 
\end{split}
\ee
Equation (\ref{e1}) can be solved to find the $Q(t)$ function, used in (\ref{Qt}), to be  

\be
\label{Qoft}
\begin{split}
Q(t) =& -\frac{a_2}{3 a_3} + \frac{1}{ 3 a_3}  \left( \frac{-q(t) + \sqrt{q(t)^2+4p(t)^3}}{2} \right)^{1/3} 
\\
&-  \frac{p(t)}{ 3 a_3}  \left(  \frac{-q(t) + \sqrt{q(t)^2+4p(t)^3}}{2} \right)^{-1/3} 
\end{split}
\ee
where
\be
&p(t) &= -a_2^2 + 3 a_1 a_3 - 3 a_3 t \\
&q(t) &=2 a_2^3 - 9 a_1 a_2 a_3 + 27 a_0 a_3^2 + 9 a_2 a_3 t.
\ee
Let us now define a new scalar field $y$ as 

\be
y = s - {\cal Q}(t)
\ee
and the new potential for $t$ and $y$ as 

\be
\label{Vty3}
 {\cal V}(t,y) =  {\cal V}(t,s)|_{s = y + {\cal Q}(t)}. 
\ee
It is then clear that since the potential ${\cal V}(t,s)$ was minimized for $s - {\cal Q}(t)=0$, 
the new potential (\ref{Vty3}) will be minimized for $y = 0$, that is  

\be
\frac{\p {\cal V}(t,y)}{\p y} \Big{|}_{y=0  }=0 . 
\ee
Thus the appropriate fields to describe a single field inflationary phase are given by $t$ and $y$. 
Here $t$ is the inflaton and $y$ will be stabilized at the origin given that $m_y^2 > H^2$. 
The scalar potential becomes 

\be 
\label{VVinfQt}
\begin{split}
{\cal V}(t,y) =& 12 m^2 M_P^2 \Big{(} 1 - 2 (y+Q(t))^2 + \frac{\zeta}{9} (y+Q(t))^4 + 2 t 
+ 2 \gamma (y+Q(t)) \Big{)}^{-2} \times 
\\
& \Big{[} (y+Q(t))^2  \lc1 - 2 (y+Q(t))^2 + \zeta (y+Q(t))^4 -2 \gamma (y+Q(t)) \rc 
\\
&+ \frac{9}{2} \frac{[t -  \gamma (y+Q(t)) 
-\frac23 (y+Q(t))^2 (3 - \zeta (y+Q(t))^2) ]^2  }{(9 - 2 \zeta (y+Q(t))^2 ) }  \Big{]}  .  
\end{split}
\ee
A plot of the potential (\ref{VVinfQt}) can be found in Fig. \ref{PLVVxt}  which 
clearly shows that we are dealing with a single field inflationary model. 
For $\langle y \rangle =0 $ the potential is further simplified to 

\be 
\label{VVinft}
\begin{split}
{\cal V}(t) \! =& 12 m^2 M_P^2 \Big{(} 1 - 2 Q(t)^2 + \frac{\zeta}{9} Q(t)^4 + 2 t + 2 \gamma Q(t) \Big{)}^{-2}  \times 
\\
& \Big{[} Q(t)^2  \lc1 - 2 Q(t)^2 + \zeta Q(t)^4 -2 \gamma Q(t) \rc 
\\
&+ \frac{9}{2} \frac{[t -  \gamma Q(t) 
-\frac23 Q(t)^2 (3 - \zeta Q(t)^2) ]^2  }{(9 - 2 \zeta Q(t)^2 ) }  \Big{]}   . 
\end{split}
\ee
From (\ref{VVinft})  we see that in the vacuum $\langle Q(t) \rangle \ne 0$, 
hence $\langle s \rangle \ne 0$ and supersymmetry is broken.

For completeness we give the  Lagrangian fot $t$ 

\be
\label{Lcompt}
e^{-1} {\cal L} = -\frac{M_P^2}{2} R - \frac12 K(t) \p t \p t - {\cal V}(t)  
\ee
where  the potential is given by (\ref{VVinft}) and the kinematic function reads 

\be
\begin{split}
K(t) &=  \frac{6 M_P^2}{(f(S, \bar S) + (T/M_P) + (\bar T/M_P))^2} \Big{|}_{(S/M_P)=Q(t), \,  (T/M_P)=t} 
\\
&= \frac{6 M_P^2}{(1+2t +2 \gamma Q(t) -2 Q(t)^2 +(\zeta/9) Q(t)^4)^2} .
\end{split}
\ee

\subsubsection{Inflationary observables}

Let us now turn to the slow-roll stage. 
In principle we would like to redefine the $t$ field such that it has a canonically normalized kinematic term, 
but due to the very involved function $K(t)$ this is not manageable.  
Thus,  the standard formulas for the slow-roll parameters can not be directly applied. 
We write the slow-roll conditions for the Lagrangian  (\ref{Lcompt}) 
in terms of the non-canonically normalized field. 
The equation of motion for the homogeneous scalar $t$ and the Friedmann equation read respectively 

\be
\begin{split}
\label{eqK}
 K(t) \ddot t + 3 H K(t) \dot t + \frac12 K'(t) \dot t^2 + {\cal V}'(t) &= 0 
\\
\frac{1}{3 M_P^2} \left( \frac12 K(t) \dot t^2 +{\cal V}(t) \right) &= H^2  
\end{split}
\ee
where $ {\cal V}'(t) = \frac{\p {\cal V}}{\p t}$, $\dot t$ refers to the cosmic time derivative and $H$ is the Hubble scale. 
Note that the $t$-field here is dimensionless. 
The non-canonical equations of motion (\ref{eqK}) give rise to generalized slow-roll parameters defined as

\be
\label{slowK}
\begin{split}
\epsilon_K &= \frac12 M_P^2 \frac{{\cal V}'^2}{K {\cal V}^2} 
\\
\eta_K &= M_P^2  \frac{{\cal V}''}{K {\cal V}} - \frac12  M_P^2   \frac{K' {\cal V}'}{K^2 {\cal V}}  . 
\end{split}
\ee
Under the conditions 

\be
\label{msl}
| \eta_K | \ll 1 \ \ , \  \ \epsilon_K \ll 1  
\ee
the equations (\ref{eqK}) approximate to 

\be
\label{sloweqK}
\dot t   \simeq - \frac{ {\cal V}'(t)}{3 H K(t) }   \  \  ,  \  \  H^2  \simeq \frac{ {\cal V}(t)}{3 M_P^2} 
\ee
and the system undergoes an inflationary phase. 
Notice that the modified slow-roll conditions (\ref{msl}) become the standard ones for $K=1$. 
Finally, the e-folds are defined as usual, but now the formula is also modified due to $K$ 

\be
\label{efolds}
{\cal N}  =  \int_{t_e}^{t_*} \frac{ K {\cal V} }{M_P^2 {\cal V}'} dt .  
\ee
Note  that $t$  refers to the $t$-field and not the cosmic time, whereas 
$t_*$ is the value of $t$ at the pivot scale and $t_e$ the value at the end of inflation.

Another important issue in inflationary models  is the effect of the additional  scalar fields. 
They have to have masses above the Hubble scale during the accelerated phase  

\be
\frac{m_{\text{scalar}}^2}{H^2} >1  
\ee
in order  that it is indeed a single field inflation. 
The fact  the fields have non-canonical kinetic terms the comparison 
between $H$ and $m_{\text{scalar}}$ is not straightforward.   
In general we can redefine the fields so as to canonically normalize the kinematic terms as 

\be
\frac{\p \hat \phi}{\p \phi}  = \sqrt{\cal B}  
\ee
such that 

\be
- \frac{1}{2}  {\cal B}(\phi) \p \phi \p \phi 
=
- \frac{1}{2}  \p \hat \phi \p \hat \phi . 
\ee

Thus the mass of the new field $\hat \phi$ is related to the original potential ${\cal V}$ 
and the kinematic function ${\cal B}$ as 

\be
\begin{split}
\label{gen}
m^2_{\hat{\phi}} &
\equiv \frac{\p}{\p \hat{\phi}}\left(\frac{\p}{\p \hat{\phi}} {\cal V}\right) 
= \frac{\p}{\p \hat{\phi}}\left(\frac{\p {\cal V}}{\p \phi} \frac{\p\phi}{\p \hat{\phi}} \right) 
\\
&= \frac{\p}{\p \hat{\phi}} \left({\cal V}'(\phi) \frac{1}{\sqrt{\cal B}} \right) 
=\frac{{\cal V}''(\phi)}{{\cal B}}\, -\frac12 \frac{{\cal V}'(\phi)}{{\cal B}^2} {\cal B}'(\phi)
\end{split}
\ee
Hence, for the strongly stabilized fields at the vacuum

\be
m^2_{\hat{\phi}} = \frac{\langle {\cal V}''(\phi) \rangle}{\langle{\cal B} \rangle} . 
\ee
From (\ref{gen})  we can find a mapping between the generalized and the standard slow roll parameters 
to verify (\ref{slowK}) 

\be
\epsilon =\frac{M^2_P}{2} \left(\frac{{\cal V}'(\hat{t})}{{\cal V}} \right)^2 
= \frac{M^2_P}{2} \frac{1}{K(t)} \left(\frac{{\cal V}'( t)}{{\cal V}} \right)^2 \equiv \epsilon_K
\ee
and 

\be
\eta =M^2_P \frac{V''(\hat t)}{{\cal V}}  
= M_P^2  \frac{{\cal V}''(t)}{K {\cal V}} - \frac12  M_P^2   \frac{K'(t) {\cal V}'(t)}{K^2 {\cal V}}\equiv \eta_K  
\ee
where we have used  ${\cal B} = K$ for the inflaton.

\begin{figure}
\centering
\begin{subfigure}{.5\textwidth}
  \centering
 \includegraphics[width=.99\linewidth]{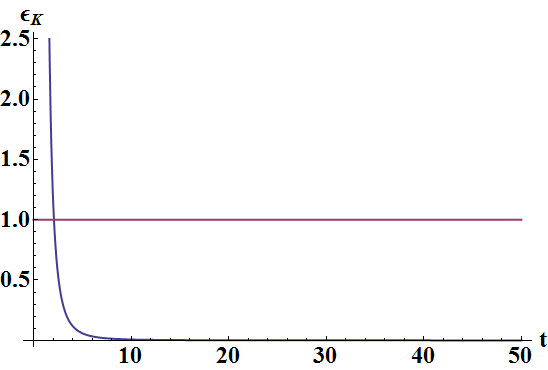}
\caption{  }
\label{epsilonplot}
\end{subfigure}%
\begin{subfigure}{.5\textwidth}
  \centering
  \includegraphics[width=.99\linewidth]{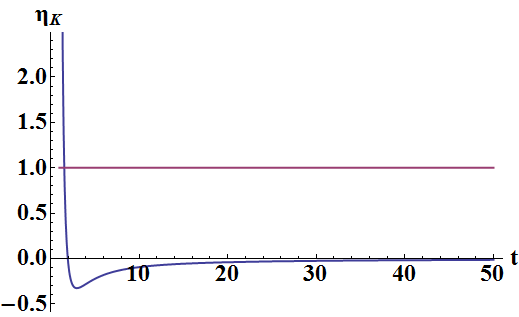}
\caption{  }
\label{masstoverV}
\end{subfigure}
\caption{The slow-roll parameters during inflation for $\gamma=1$ and $\zeta=8$ case. 
The $\epsilon_K$- and $\eta_K$- parameter during inflation with respect to the $t$-field value (left and right panel respectively). }
\label{fig:test}
\end{figure}

\begin{figure}
\centering
\begin{subfigure}{.33\textwidth}
  \centering
  \includegraphics[width=.99\linewidth]{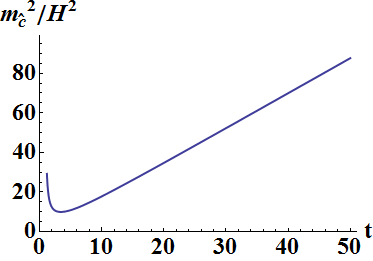}
\caption{  }
\label{masscoverH}
\end{subfigure}%
\begin{subfigure}{.33\textwidth}
  \centering
  \includegraphics[width=.99\linewidth]{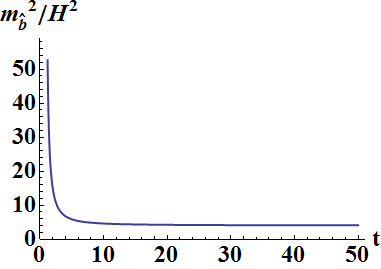}
 \caption{   }
\label{massdoverH}
\end{subfigure}
\begin{subfigure}{.33\textwidth}
  \centering
  \includegraphics[width=.99\linewidth]{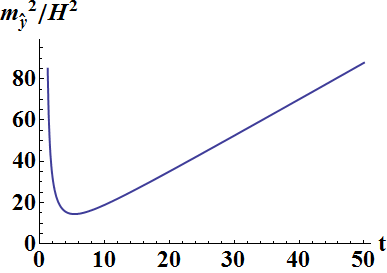}
 \caption{ }
\label{massxoverH}
\end{subfigure}
\caption{Plots of the masses squared of the stabilized scalars $\hat c$, $\hat b$ and $\hat y$ over $H^2$ during inflation for $\gamma=1$ and $\zeta=8$ case (from left to right panel). }
\label{fig:test}
\end{figure}

Plots of $m_{\hat j}^2 / H^2$ where $j = c, b, y$ 
can be found in  Fig.  \ref{masscoverH}, Fig.  \ref{massdoverH} and Fig. \ref{massxoverH}.  
The only non-diagonal  term in the scalar  mass matrix is the $t-y$ term, 
which is of order $m^2_{t-s} \sim {\cal O}(\frac{m_t^2 }{10}) \ll m^2_y $ and thus negligible. 
For the   $t$-field  a plot of  the $\eta_K$- and $\epsilon_K$-parameter, can be found in Fig. \ref{masstoverV} 
and in Fig. \ref{epsilonplot} respectively.

The canonically normalized $t$-field (inflaton) 
mass in the vacuum, $\p^2 {\cal V} / \p \hat t \p \hat t $,  is  found to be  
\begin{align}
m^2_{inf}&=
m^2\left(-9259827984  + 114906047256 s_0^2 - 98720480160 s_0^4 + 
 22001657400 s_0^6\right)^{-1}\times \nonumber \\
 &\left(64s_0^{-6} + 1288 s_0^{-4} + 26244 s_0^{-3} + 74245 s_0^{-2} + 26244 s_0^{-1} - 63649 - 
 52488 s_0\right. \nonumber \\
 &\left.- 59680 s_0^2 + 2048 s_0^4\right)^2  
 \end{align} 
and it depends on $s_0$ directly and also via the $\zeta_0$ and the $\gamma_0$ parameters. 

For a $\zeta$-parameter of order 10, $s_0$ is of order a half, 
and the gravitino and inflaton  mass turn out to be  

\be
m^2_{3/2} \simeq  m^2    \, , ~~~m^2_{inf}\simeq \frac{1}{4} m^2\, .
\ee
Hence these models  imply a relation of the inflaton mass to the gravitino mass $m_{3/2}$ 
via the new scale $m$.


As an illustrative example  we further study the case
\be
\zeta =8 \ \ , \ \ \gamma =1   .  
\ee 
For these parameters $\zeta$ and $\gamma$ it is straightforward to find  
that the value of $t$ at the end of inflation ($\epsilon_K \simeq 1$) 
is $t_e=2.1$. 
For an approximate number of $N(t_*) \simeq 55$  e-folds (\ref{efolds}), we find the value $t_* = 50$. 
Therefore the  tensor-to-scalar ratio $r$ is 
\be
r = 16 \epsilon(t_*) \simeq 8 \times 10^{-3}   
\ee
at the pivot scale $t_*$. 
Thus the model predicts very small amount of gravitational waves. 
There is another way to understand this. 
For large $t$  values ($t>20$)  it is consistent to approximate 

\be
\langle y \rangle  \simeq s_* \ll 1 . 
\ee
This leads to the inflationary effective Lagrangian for $t$ 

\be
e^{-1} {\cal L} = - \frac{M_P^2}{2} R 
- \frac{3 M_P^2}{(1+ 2 s_* + 2 t )^2} \p t \p t 
- 6 m^2 M_P^2 \frac{(t- s_*)^2}{(1+ 2 s_* + 2 t )^2}  . 
\ee
After a redefinition of the $t$ field 

\be
t = \frac12 \left( e^{\sqrt \frac23 \varphi/M_P} -1 - 2 s_* \right)
\ee
we have

\be
\label{larget}
e^{-1} {\cal L} = - \frac{M_P^2}{2} R 
- \frac12 \p \varphi \p \varphi 
- \frac32 m^2 M_P^2 \left( 1  - (1 + 4 s_*) e^{- \sqrt \frac23 \varphi/M_P}  \right)^2 . 
\ee
From the number $\frac23$ in the exponential we see that the model 
is similar to the original Starobinsky model 
and will give rise to the similar amount of gravitational waves 
(see also \cite{Kallosh:2013hoa}), 
which is favored by the Planck collaboration data \cite{Planck}. 
Note that for the  Starobinsky model in supergravity $s_* = 0$ always.

\begin{figure}
\centering{
\includegraphics[width=.5\linewidth]{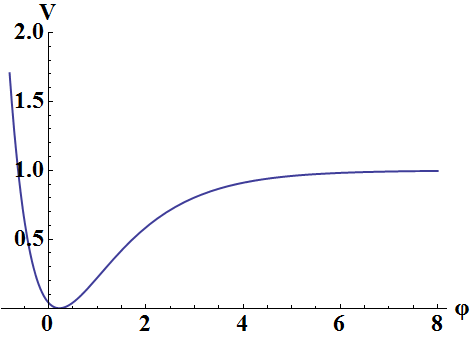}}
\label{new-single-large}
\caption{A plot of the scalar potential in (\ref{larget}) for the inflaton field during inflation. }
\end{figure}

\subsection{Vacuum structure for $f_{\rm v} = \frac{\beta}{M_P^3} \left( {\cal S} \bar {\cal S}^2 +\bar {\cal S} {\cal S}^2\right)$}

A second example of higher curvature supergravity that drives single field inflation and breaks supersymmetry 
is described by the model 

\be
\label{RRRR2}
f ({\cal R} , \bar {\cal R})= 1 + \beta \, \frac{ {\cal R} \bar{\cal R}^2 + {\cal R}^2\bar {\cal R }}{m^3}
- 2  \frac{ {\cal R} \bar {\cal R}}{m^2}  
+ \frac19 \zeta \, \frac{{\cal R}^2 \bar {\cal R}^2}{m^4}  
\ee
The second term in (\ref{RRRR2}) source the R-symmetry breaking. The equivalent K\"ahler and super potential  read 

\be
{\cal K} = - 3 M_P^2 \, \text{ln} \lc 1   +  \frac{{\cal T}  + \bar {\cal T}}{ M_P}  
+  \beta \frac{{\cal S}^2 \bar {\cal S} + {\cal S} \bar {\cal S}^2}{ M_P^3} 
- 2  \frac{{\cal S} \bar {\cal S}}{M_P^2} 
+ \frac19 \zeta \, \frac{{\cal S}^2 \bar {\cal S}^2}{M_P^4}  \rc  
\ee
\be
{\cal W} = 6 m \, {\cal T} \, {\cal S} 
\ee
that yield the scalar potential

\be 
\label{VV2}
\begin{split}
{\cal V}(t,b, s,c) =& 12 m^2 M_P^2 \Big{(} 1 - 2 (s^2 + c^2 ) + \frac{\zeta}{9} (s^2 + c^2 )^2 + 2 t + 2 \beta s (s^2+c^2) \Big{)}^{-2} \times 
\\
& \Big{[} (s^2 + c^2) \lc1 - 2(s^2+c^2) + \zeta (s^2 + c^2)^2 +6 \beta s (s^2 + c^2) \rc  
\\
& - \frac{[t -2(s^2 +c^2) +5 \beta s c^2 +5 \beta s^3 + \frac23 \zeta (s^2 +c^2)^2 ]^2 
+  (b - \beta c^3 - \beta c s^2)^2 }{(4 \beta s -2 + \frac49 \zeta (s^2 + c^2))}  \Big{]}    
\end{split}
\ee
where $T/M_P=t+ib$ and $S/M_P=s+ic$. 
This potential generically has two classes of vacuum solutions:
the supersymmetric Minkowski vacuum 
\be
\label{v02}
\langle  T \rangle = \langle  S \rangle = 0 
\ee
and a new class of vacua 
\be
\label{vBR2}
\begin{split}
\langle  T \rangle &= M_P \langle  t \rangle + i M_P \langle b \rangle = M_P \, t_0  
\\
\langle  S \rangle &=  M_P \langle  s \rangle + i M_P \langle c \rangle   = M_P \, s_0  
\end{split}
\ee
which break supersymmetry with vanishing or small positive (or negative) vacuum energy depending on the parameters $\zeta$ and $\beta$. 

The potential involves the $(b - \beta c^3 - \beta c s^2)^2$ term and even powers of the filed $c$ which imply that 
\be
\langle c \rangle = \langle b \rangle = 0 .  
\ee
This has been also  verified numerically. 
Considering that the $b$ and the $c$ field are stabilized at the origin the potential reads

\be 
\label{VVapp2}
\begin{split}
{\cal V}(t,s) =& 12 m^2 M_P^2 \Big{(} 1 - 2 (s^2 ) + \frac{\zeta}{9} (s^2  )^2 + 2 t + 2 \beta s (s^2) \Big{)}^{-2} \times 
\\
& \Big{[} s^2 \lc1 - 2 s^2 + \zeta s^4 +6 \beta s^3 \rc  
 - \frac{[t -2 s^2 +5 \beta s^3 + \frac23 \zeta s^4 ]^2 }{(4 \beta s -2 + \frac49 \zeta s^2)}  \Big{]}\,.    
\end{split}
\ee
Following the same reasoning as in the previous sections we find that the supersymmetry breaking vacua are Minkowski ones for the parameter space
\be
\zeta &=& \frac{3 - 2 s_0^2}{ s_0^4} 
\\
\beta &=& \frac{2 s_0^2 -2}{3 s_0^3}  \,.
\ee
The gravitino mass at this vacuum has the value

\be
\label{grmass2}
m^2_{3/2} = \langle e^{\cal K} |{\cal W}|^2 \rangle = m^2  \frac{729 s_0^2}{8 (3- s_0^2)^3}
\ee 
The above formula (\ref{grmass2}) offers a non-trivial bound on the possible values of $s_0$ 
\be
|s_0| < \sqrt 3 
\ee
such that the gravitino mass squared is positive. 
This also implies $\zeta > -0.1 $ and $\beta < 1.48$. 
Moreover note that even though supersymmetry is broken and the vacuum energy is vanishing the gravitino 
mass is not specified until the parameter $s_0$ is chosen.

\begin{figure}[tbp] 
\centering{
\includegraphics[scale=0.52]{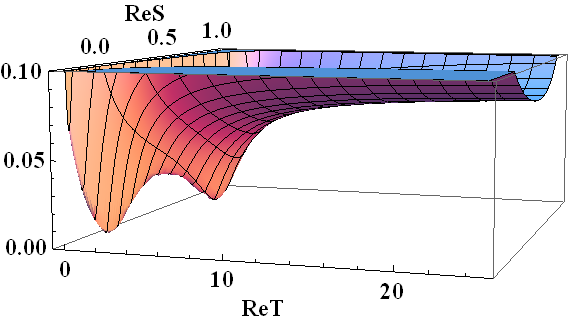}}
\caption{The shape of the scalar potential (\ref{VV2}) with $\zeta=8$ 
and $\beta = -\frac{2 \sqrt 2 }{3 }$ in units of $12 m^2 M_P^2$. 
The scalar fields $T$ and $S$ are in $M_P$ units.  
A plateau similar to the Starobinsky model can be seen for large Re$T$ values. 
The inflationary trajectory terminates at the SUSY breaking vacuum where Re$S \simeq 0.7$.}
\label{linSVV2}
\end{figure}

To illustrate the features of the potential let as choose specific vaues for the $\zeta$ and $\beta$ parameters
\be \label{zb}
\zeta = 8\,,\quad   
\beta = -\frac{2 \sqrt 2 }{3 }. 
\ee
Apart from the supersymmetric vacuum at the origin there is also the supersymmetry breaking one that lies at
\be
\begin{split}
\langle  T \rangle &= M_P t_0 =  \frac{4}{3} M_P  
\\
\langle  S \rangle &= M_P s_0   =  \frac{1}{\sqrt 2} M_P  .  
\end{split}
\ee
All the vacua have vanishing vacuum energy, see Fig. \ref{linSVV2}. 
The vacuum expectation values of these fields are here marginally transPlanckian. 
A consistency check for the kinematic terms shows that they are positive definite around this vacuum 
\be
\left( f + \frac{T  }{M_P} + \frac{\bar T}{  M_P} \right) f_{S \bar S} < 0\,.  
\ee 
The gravitino mass is found to be
\be
m^2_{3/2} = \langle e^{\cal K} |{\cal W}|^2 \rangle \simeq \frac{73}{25}  m^2   
\ee 
and $\langle D_T {\cal W} \rangle \approx \langle D_S {\cal W} \rangle \approx  \, M \, M_P$.

We note that one can obtain de Sitter or anti-de Sitter vacua  for $\zeta=8$ and 
\be
\beta \gtrsim  -\frac{2 \sqrt 2 }{3 }\,, \quad\quad
\beta \lesssim  -\frac{2 \sqrt 2 }{3 }
\ee
respectively.

\subsection{Inflationary properties for 
$f_{\rm v} = \frac{\beta}{M_P^3} \left( {\cal S} \bar {\cal S}^2 +\bar {\cal S} {\cal S}^2\right)$}

\begin{figure}
\centering
\begin{subfigure}{.48\textwidth}
  \centering
  \includegraphics[width=.99\linewidth]{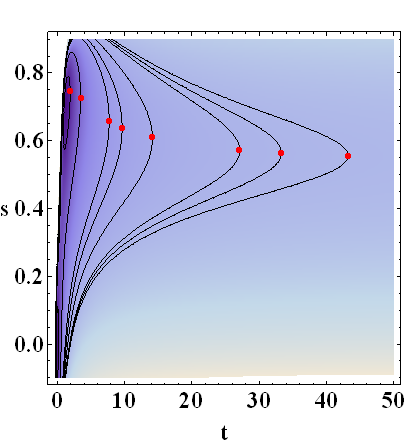}
  \caption{}
  \label{PLVVinf2}
\end{subfigure}%
\begin{subfigure}{.5\textwidth}
  \centering
  \includegraphics[width=0.99\linewidth]{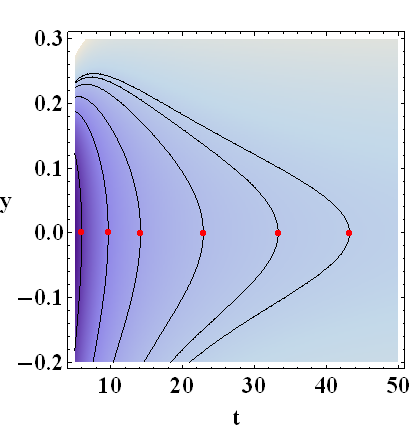}
  \caption{}
  \label{PLVVxt2}
\end{subfigure}
\caption{The trajectory in field space Re$T$ - Re$S$ for $\beta=-\frac{2 \sqrt 2 }{3 }$ and $\zeta=8$ 
for the original fields in the scalar potential (\ref{VV2}) (left) and for the redefined fields in (\ref{VVapp3}) (right). }
\label{fig:test}
\end{figure}

In order to find the single-field inflationary trajectory we follow the procedure discussed earlier. We define the new field variable $y = s - Q(t)$ such that the new potential ${\cal V}(t,y)$ has a valey along  the $y=0$ direction
\be
\frac{\p {\cal V}(t,y)}{\p y} \Big{|}_{y=0} =0. 
\ee 
The potential (\ref{VVapp2}) now reads

\be 
\label{VVapp3}
\begin{split}
{\cal V}(t,y) =& 12 m^2 M_P^2 \Big{(} 1 - 2 (y+Q(t))^2  + \frac{\zeta}{9} (y+Q(t))^4 + 2 t 
+ 2 \beta (y+Q(t))^3 ) \Big{)}^{-2} \times 
\\
& \Big{[} (y+Q(t))^2 \lc1 - 2 (y+Q(t))^2 + \zeta (y+Q(t))^4 +6 \beta (y+Q(t))^3 \rc  
\\
& - \frac{[t -2 (y+Q(t))^2 +5 \beta (y+Q(t))^3 + \frac23 \zeta (y+Q(t))^4 ]^2 }{(4 \beta (y+Q(t)) -2 
+ \frac49 \zeta (y+Q(t))^2)}  \Big{]}  .  
\end{split}
\ee
For plausible values for the parameters $\zeta$ and $\beta$ the mass of the $y$ field can be larger than the Hubble scale during inflation and stabilized at $\langle y \rangle =0$. Hence the potential effectively depends on a single field, the $t$, whose Lagrangian reads

\be
\label{Lcompt6}
e^{-1} {\cal L} = -\frac12 M_P^2 R - \frac12 K(t) \p t \p t - {\cal V}(t)\,.  
\ee
The above kinematic function is given by the expression

\be
\begin{split}
K(t) &=  \frac{6  M_P^2}{(f + T/M_P + \bar T/M_P)^2}  \Big{|}_{(S/M_P)=Q(t), \,  (T/M_P)=t} 
\\
&= \frac{6 M_P^2}{(1+2t +2 \beta Q(t)^3 -2 Q(t)^2 +(\zeta/9) Q(t)^4)^2}.
\end{split}
\ee
In order to turn the non-canonical normalized kinematic term into a canonical form a field redefinition is required, however this cannot be done in straightforward manner.
Nevertheless, the slow roll parameters and the number of e-folds can be calculated using the formulas (\ref{slowK}) and (\ref{efolds}) respectively.

\begin{figure}
\centering
\begin{subfigure}{.33\textwidth}
  \centering
  \includegraphics[width=.99\linewidth]{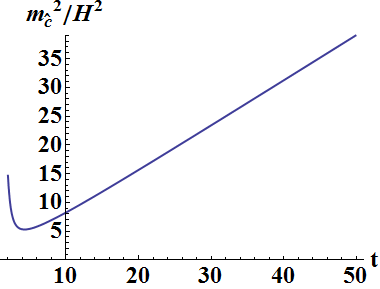}
\caption{  }
\label{masscoverH2}
\end{subfigure}%
\begin{subfigure}{.33\textwidth}
  \centering
  \includegraphics[width=.99\linewidth]{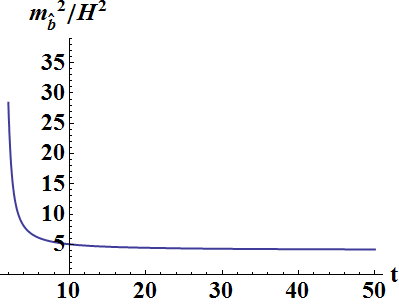}
 \caption{   }
\label{massdoverH2}
\end{subfigure}
\begin{subfigure}{.33\textwidth}
  \centering
  \includegraphics[width=.99\linewidth]{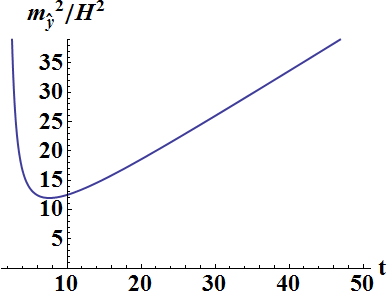}
 \caption{ }
\label{massxoverH2}
\end{subfigure}
\caption{Plots of the masses of the stabilized scalars over $H^2$ during inflation for $\zeta=8$ and $\beta=-\frac{2 \sqrt 2 }{3 }$. 
From left to right: the mass of the $\hat c$ field over $H^2$,   the $\hat d$ field over $H^2$  and 
the $\hat y$ field over $H^2$. }
\label{fig:test}
\end{figure}

Let us see the inflationary features and observables for the specific model (\ref{zb}). 
The trajectory, $y=0$, is specified by the equation 
\be
\frac{\p {\cal V}(t,s)}{\p s}\Big{|}_{s={Q}(t)} = 0 
\ee
which can be solved by using the approximation
\be \label{ts}
t = b_0 + b_{-1} (s - s_i)^{-1} + b_{-2} (s - s_i)^{-2} + b_{-3} (s - s_i)^{-3}.  
\ee
Here $s_i \sim 0.5$ is the $s$ field value in the beginning of inflation,  
$t\gg 1$, and the expansion coefficients are found by a numerical fit to be
\be
\begin{split}
b_0 &= -7.40686 
\\
b_{-1} &= 3.07583 
\\
b_{-2} &= 0.16067 
\\
b_{-3} &= 0.00839\,.
\end{split}
\ee
The approximation (\ref{ts}) allows us to find the form of $Q(t)$ in the same fashion as in (\ref{Qoft}), and hence, with great precision the inflationary potential and trajectory.
Equipotential contours for the scalar potential ${\cal V}(t, s)$ and ${\cal V}(t, y)$ are depicted in Fig. \ref{PLVVxt2}. 
This is the field trajectory because the scalar fields $\hat b, \hat c$ and $\hat y$ 
have masses larger than the Hubble scale during inflation and get stabilized at their vacuum, see the plots in Fig. \ref{masscoverH2}, Fig. \ref{massdoverH2} and Fig. \ref{massxoverH2}. 
In this example, as in the previous one, the field trajectory terminates in the supersymmetry breaking vacuum.

The mass of the canonically normalized t-field (inflaton) in the vacuum is given by
\begin{eqnarray}
m_{\hat t}^2= \frac{\p^2 {\cal V}}{\p \hat t \p \hat t} = m^2
\frac{(192 - 576 s_0 + 88 s_0^2 + 252 s_0^3 - 63 s_0^4 + 324 s_0^5 - 405 s_0^6)^2}{944784 s_0^6 - 472392 s_0^8 + 17496 s_0^{12}} . 
\end{eqnarray}
For the values (\ref{zb})  we take the mass squareds  for the $\hat t$-field and the 
gravitino in the vacuum 
\be
m_{3/2}^2 \simeq 3 m^2 \  \  ,  \  \  m^2_{\hat t} \simeq \frac{1}{10 } m^2 . 
\ee

\begin{figure}
\centering
\begin{subfigure}{.5\textwidth}
  \centering
 \includegraphics[width=.99\linewidth]{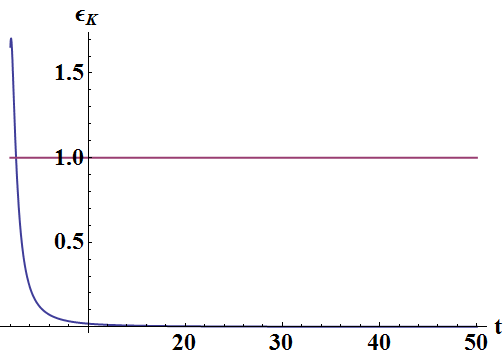}
\caption{  }
\label{epsilonplot2}
\end{subfigure}%
\begin{subfigure}{.5\textwidth}
  \centering
  \includegraphics[width=.99\linewidth]{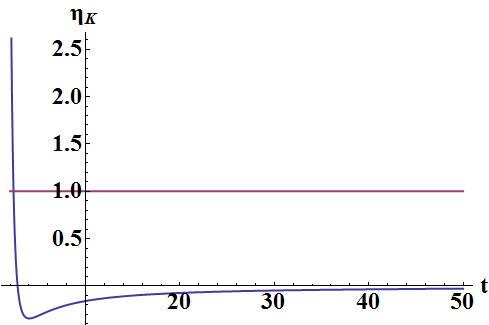}
\caption{  }
\label{masstoverV2}
\end{subfigure}
\caption{The slow-roll parameters during inflation for $\beta=-\frac{2 \sqrt 2 }{3 }$ and $\zeta=8$. 
The $\epsilon_K$-parameter during inflation for $t$ (left panel). 
The $\eta_K$-parameter during inflation for $t$ (right panel).}
\label{fig:test}
\end{figure}

Inflation is driven for slow roll parameters $\epsilon_K, |\eta_K|$. 
Their values depend on the $t$ field and the corresponding 
plots can be found in Fig.  \ref{epsilonplot2}, \ref{masstoverV2} . 
Inflation ends when $\epsilon_K= 1$ which is found to take place 
at $t_e=2.6$. For an approximate number of $N \simeq 55$ e-folds, 
inflation starts from $t_i = 80$ which we can identify with the pivot scale $t_*$. 
For this field value the tensor-to-scalar ratio $r$ reads
\be
r_* = 16 \epsilon(t_*) \simeq 3.2 \times 10^{-3}.   
\ee
Hence, this model predicts a very small amount of gravitational waves in accordance with the results of the Planck collaboration \cite{Planck}.

\section{Conclusions}

In this work we have exploited the properties of pure supergravity in order to 
show that it may account for both 
the inflationary phase of our universe and the  breakdown of supersymmetry.  
Our investigation was carried out in the framework of higher curvature old-minimal supergravity.  
This higher curvature theory is equivalent to 
standard supergravity coupled to two chiral supermultiplets. 

We have illustrated how the well known R-symmetric models can not  give a stable inflationary regime 
with supersymmetry breaking in the vacuum.
This is expressed by the fact that the $\zeta$-parameter, in front of the higher curvature ${\cal R}^2\bar{\cal R}^2/m^4$ term, can take no values such that both the aforementioned phenomena  to be accomplished: small values for the $\zeta$-parameter yield supersymmetry breaking vacua but such small value cannot implement the inflationary phase.
 
We have shown that when one departs from exact R-symmetric theories, by taking into account leading order R-violating terms, stable inflationary trajectories that end up in consistent  supersymmetry breaking vacua appear. These non-supersymmetric vacua can be tuned to be Minkowski, therefore no extra terms are required in order to set the cosmological constant to zero.  In these models the inflaton and the gravitino masses are of the same order of magnitude that is much higher than the electroweak scale. Furthermore, the corresponding Polonyi field that implements the supersymmetry breakdown can be identified as part of the inflaton field. Thereby by construction these models do not suffer from any cosmological problem associated with a light gravitino or Polonyi field. 
Despite the high scale of supersymmetry breaking a low energy soft supersymmetry breaking mass term pattern can be achieved with a moderate degree of fine-tuning in the transmission scenario \cite{Hindawi:1996qi}. 

Summarizing, we found that a wide class of a higher curvature gravitational models can account for both the cosmic inflationary phase and the supersymmetry breakdown.  The inflationary predictions are  similar to those of the Starobinsky model. 
The analysis of the mediation of the breaking to the low energy physics is left for future work. 
We stress that the models we discussed here are of pure supergravity origin, 
and no additional sector is invoked. Therefore these models provide a minimalistic and unifying description of the inflationary phase and the supersymmetry breakdown.

\section*{Acknowledgements}

I.D. would like to thank Masaryk university for hospitality during the preparation of the present work. 
The work of F.F. and R.v.U. is supported by the Grant agency of the Czech republic under the grant P201/12/G028. 
The research of A.K. was implemented under the Aristeia II Action of the Operational Programme Education and Lifelong Learning and is co-funded by the European Social Fund (ESF) and National Resources. 
A.K. is also partially supported by European Union's Seventh Framework Programme (FP7/2007-2013) 
under REA grant agreement n. 329083. 
A.R. is supported by the Swiss National Science Foundation (SNSF), 
project ``The non-Gaussian Universe" (project number: 200021140236).

\end{document}